\documentclass[aps,prb,twocolumn,superscriptaddress]{revtex4-2}
\usepackage{float}
\usepackage{amsmath,array,amssymb}
\usepackage{graphicx}
\usepackage{siunitx,bigints}
\usepackage{multirow}
\usepackage{amsfonts}
\usepackage{setspace}
\usepackage[infoshow,debugshow]{tabularx}
\usepackage{braket}

\usepackage{hyphenat}
\usepackage[english]{babel}
\usepackage[english]{babel}

\usepackage{lineno}
\usepackage{color}
\usepackage{multirow}
\usepackage{booktabs}
\usepackage{threeparttable}
\usepackage{adjustbox}
\usepackage{xr-hyper}
\makeatletter
\usepackage{filecontents}
\newcommand*{\addFileDependency}[1]{
\typeout{(#1)}
\@addtofilelist{#1}
\IfFileExists{#1}{}{\typeout{No file #1.}}
}\makeatother
\newcommand*{\myexternaldocument}[1]{%
\externaldocument{YbYSO_zSM}%
\addFileDependency{YbYSO_zSM}%
\addFileDependency{YbYSO_zSM}%
}

\myexternaldocument{YbYSO_zSM}
\usepackage{hyperref}
\begin{document}
	
	\title{Optical coherence and spin population dynamics in $^{171}$Yb$^{3+}$:Y$_2$SiO$_5$ single crystals}
	
	\author{Federico Chiossi} 
 \email{federico.chiossi@chimieparistech.psl.eu}	
 \affiliation{Chimie ParisTech, PSL University, CNRS, Institut de Recherche de Chimie, Paris, 75005, France}

\author{Eloïse Lafitte-Houssat}	
\affiliation{Chimie ParisTech, PSL University, CNRS, Institut de Recherche de Chimie, Paris, 75005, France}
 \affiliation{Thales Research and Technology, 91767 Palaiseau, France}
\author{Alban Ferrier}	
\affiliation{Chimie ParisTech, PSL University, CNRS, Institut de Recherche de Chimie, Paris, 75005, France}
\affiliation{Faculté des Sciences et Ingénierie, Sorbonne Université, UFR 933, 75005 Paris, France}

\author{Sacha Welinski}
\author{Lo\"ic Morvan}

 \affiliation{Thales Research and Technology, 91767 Palaiseau, France}

\author{Perrine Berger}
\affiliation{Thales Research and Technology, 91767 Palaiseau, France}
 
\author{Diana Serrano}
\affiliation{Chimie ParisTech, PSL University, CNRS, Institut de Recherche de Chimie, Paris, 75005, France}
\author{Mikael Afzelius}
 \affiliation{Department of Applied Physics, University of Geneva, Geneva 4, Switzerland}
 
\author{Philippe Goldner}
\email{philippe.goldner@chimieparistech.psl.eu}	
 \affiliation{Chimie ParisTech, PSL University, CNRS, Institut de Recherche de Chimie, Paris, 75005, France}

\begin{abstract}
\noindent
$^{171}$Yb$^{3+}$-doped Y$_2$SiO$_5$ (YSO) crystals are a promising platform for optical quantum memories in  long-distance quantum communications. The relevance of this material lies in $^{171}$Yb long optical and spin coherence times, along with a large hyperfine splitting, enabling long quantum storage over large bandwidths. Mechanisms affecting the optical decoherence are however not precisely known, especially since low-temperature measurements have so far focused on the 2 to 4 K range.  
In this work, we performed two- and three-pulse photon echoes and spectral hole burning to determine optical homogeneous linewidths  in two $^{171}$Yb:YSO crystals doped at 2 and 10\,ppm. Experiments were performed in the 40\,mK to 18\,K temperature range, leading to linewidths between 320\,Hz, among the narrowest reported for rare-earth ions, and several MHz. 
Our results show that above $\sim$6\,K the homogeneous linewidth $\Gamma_h$  is mainly due to an elastic two-phonon process which results in a slow broadening with temperature, $\Gamma_h$ reaching only 25\,kHz at 10\,K.  At lower temperatures, interactions with $^{89}$Y nuclear spin-flips, paramagnetic defects or impurities, and also Yb-Yb interactions for the higher concentrated crystal, are likely the main limiting factor to $\Gamma_h$. In particular, we conclude that the direct effect of spin and optical excited state lifetime is a minor contribution to optical decoherence  in the whole temperature range studied.  Our results indicate possible paths and regimes for further decreasing of homogeneous linewidths or maintaining narrow lines at higher $^{171}$Yb concentration. 
		
\end{abstract}
	
\maketitle
	
\modulolinenumbers[2]
\section{Introduction}
The interest in rare-earth ions doped crystals (REC) for applications in quantum technologies has intensified in recent years\,\cite{Thiel2011,kunkel2018,Kinos2021}. In particular, the presence of infrared transitions in the low-loss band of fiber-optical telecommunication having millisecond-long coherence time\,\cite{Bottger2009} makes REC an ideal platform for the realization of quantum repeaters\,\cite{Afzelius2015}. These devices are of fundamental importance to overcome the limitation in quantum information transmission over long distances. 

The atomic-frequency comb (AFC) quantum memory, proposed in Ref.\,\onlinecite{afzelius2009} and realized in several REC\,\cite{lago2021telecom,laplane2015multiplexed}, is one of the most reliable approaches to develop quantum repeaters, enabling the storing of several temporal modes with also the possibility of multiplexing both in frequency and in space\,\cite{ortu2022multimode}. 
Yttrium orthosilicate crystals doped with ytterbium-171 ($^{171}$Yb:YSO) have been demonstrated to be a good candidate for broadband AFC storage\,\cite{Businger2020, Moritz_thesis}. By exploiting its transitions at 979\,nm, a quantum memory with a capability of 1250 temporal modes, a storing time of 25\,$\mu$s and a bandwidth of 100\,MHz was recently achieved\,\cite{businger2022non}. 

The exceptional properties exhibited by the $^{171}$Yb$^{3+}$ Kramers ions ($I=1/2$, $S=1/2$) in YSO crystal at zero magnetic field are due to a specific mixing of the nuclear and electron wavefunctions. It indeed gives rise to zero first-order Zeeman (ZEFOZ) transitions, also known as clock transitions in atomic physics\,\cite{Ortu2018}. In other words, under this peculiar ZEFOZ point condition, the $^{171}$Yb ions in any energy level have their magnetic moments quenched and, like non-Kramer's ions, are thus weakly coupled to the environment through magnetic interactions\,\cite{Welinski2016, Nicolas2022coherent}. Their optical and spin transitions may achieve long coherence times, up to about 1 and 10\,ms respectively \cite{Nicolas2022coherent,lafitte2022optical}, that are also related to the time during which quantum information can possibly be stored. Unlike non-Kramer's ions, however, $^{171}$Yb ions maintain their electron spin degree of freedom and the resulting large hyperfine splitting is the key to attaining quantum memories with an operational bandwidth of hundreds of MHz through the AFC protocol\,\cite{businger2022non,Moritz_thesis}. 
Long coherence times can also be obtained for Kramers ions in more technically challenging environments such as ultra-low temperatures and concentrations \cite{ledantecTwentythreeMillisecondElectron2021}or magnetic fields of several Teslas \cite{Bottger2006,rancicCoherenceTimeSecond2018}. In these cases, it may be difficult however to simultaneously observe long-lived optical and spin $T_2$, as well as high enough absorption and efficient spectral tailoring. 

$^{171}$Yb has also been used in YVO$_4$ and LiNbO$_3$ crystals for spectroscopic investigations \cite{chiossi2022photon,kindemCharacterization171Yb3YVO42018}, to demonstrate single ion detection \cite{xiaTunableMicrocavitiesCoupled2022,Kindem2020} and controlled interactions with host spins \cite{ruskucNuclearSpinwaveQuantum2022}, and investigate microwave to optical transduction \cite{bartholomewOnchipCoherentMicrowavetooptical2020}. 

Despite several spectroscopic studies devoted to $^{171}$Yb:YSO, its optical and spin coherence properties have been explored mostly in a small temperature range, 2 to 4 K,  and for the optical transitions, the underlying dephasing mechanisms are not fully understood yet\,\cite{Welinski2016,Welinski2020,Nicolas2022coherent, lafitte-houssatOpticalHomogeneousInhomogeneous2022}.  
An optical coherence time of 300\,$\mu$s was measured in a $^{171}$Yb:YSO crystal doped at 10\,ppm. It could be extended, by optically polarizing the $^{171}$Yb spin,  to 800\,$\mu$s \cite{Welinski2020}. After the Yb-Yb flip-flop was singled out as one of the main decoherence sources, crystals with lower Yb concentrations, equal to 5\,ppm and 2\,ppm, were grown, and optical coherence times of $610\pm50\,\mu$s for the former and $1.05\pm0.13\,$ to $1.25\pm 0.08$ ms for the latter were measured\,\cite{Nicolas2022coherent,lafitte-houssatOpticalHomogeneousInhomogeneous2022}. 



In this paper, the main dephasing factors of the $^{171}$Yb optical transitions are investigated at the zero-field ZEFOZ point in two YSO crystals doped with 10\,ppm and 2\,ppm. For this purpose, we first determined  spin lifetimes on relevant time scales from 40\,mK to 10\,K by measuring and modeling spectral hole and three-pulse photon echo (3PE) decays. The optical coherence and spectral diffusion parameters were then measured between 40\,mK and 18\,K by photon echo (PE) and spectral hole burning (SHB) experiments. From these results, we could identify different contributions to optical decoherence, such as direct (lifetime) and indirect (magnetic noise) effects of spin relaxation, or two-phonon scattering. They depend on $^{171}$Yb concentration and temperature range, and explain most of the observed values, except at the lower concentration and temperature for which a residual dephasing is observed.


\section{Experimental}

Measurements were performed using a setup similar to the one described in Ref. \onlinecite{chiossi2022photon}. Briefly, samples were cooled down to 40\,mK in a dilution refrigerator (BlueFors SD) and optically excited using a tunable diode laser (Toptica DL Pro, linewidth $\leq$100\,kHz). Light pulses and frequency scans  were created using acousto-optic modulators (AOM, AA Opto Electronic) driven by an arbitrary waveform generator (Keysight M3201A) and signals gated by an AOM were detected by an avalanche photodiode (Hamamatsu c5460). More information on SHB and PE experimental procedures can be found in the Appendix and Supplementary Material (SM) sec.\,\ref{sec:setup}.  

\begin{figure}[h!]
	\includegraphics[width=1\linewidth]{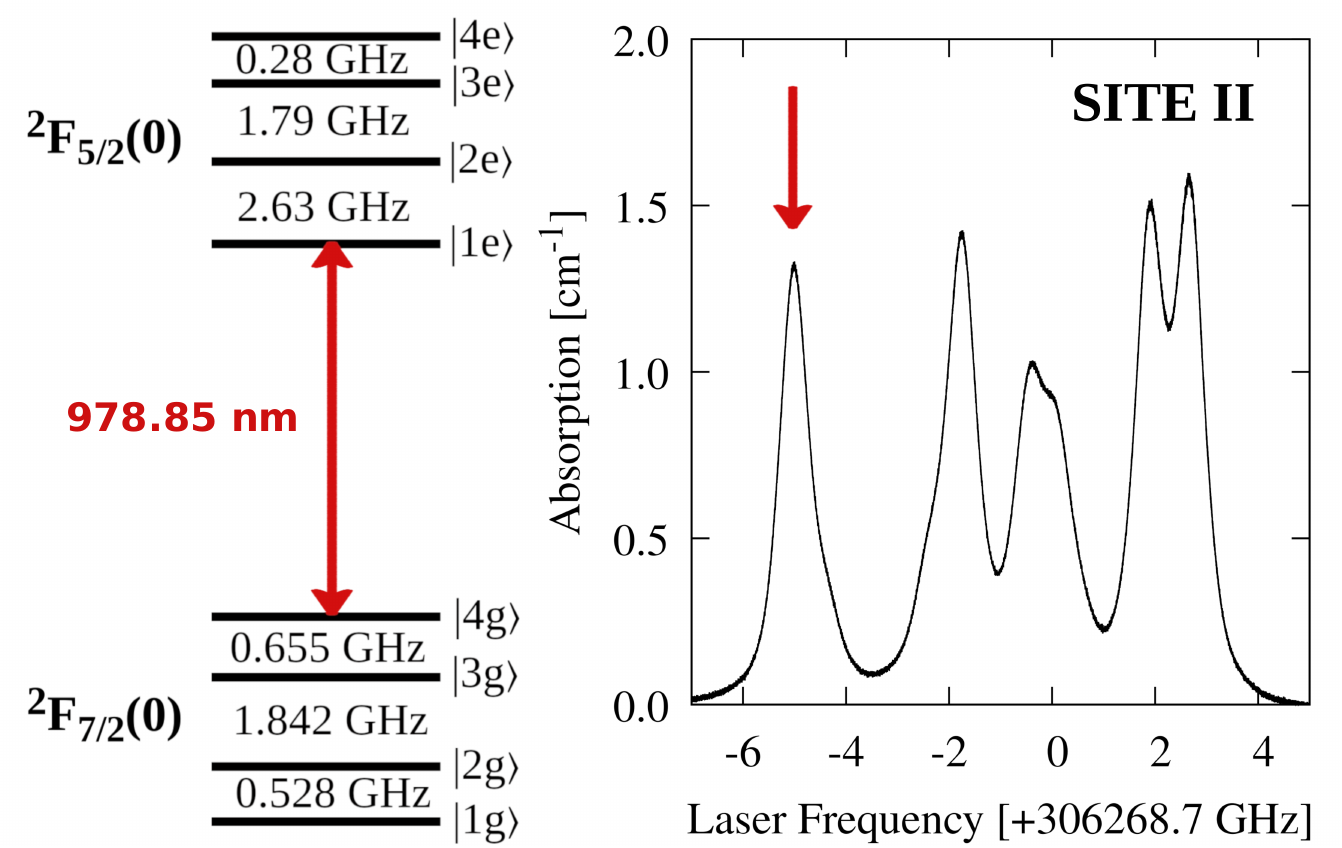}
	\caption{(Left) Hyperfine splitting of the ground and optical levels of $^{171}$Yb$^{3+}$ in site II of the YSO matrix. (Right) Absorption spectrum recorded at 4.5\,K of the $^{171}$Yb:YSO crystal (site II) doped at 10\,ppm. The red arrows indicate the transition probed in this work. }
	\label{fig:absorption}
\end{figure}

The YSO matrix has a monoclinic structure (C$_2$/c space group, \# 15) with two non-equivalent sites of $C_1$ point symmetry where ytterbium ions can substitute yttrium ones. The samples were grown by the Czochralski method using $^{171}$Yb$_2$O$_3$ raw material enriched at 95\%. Samples were cut along the principal dielectrics axes $D_1,D_2,b$\,\cite{Wyon1992} with dimensions of $5\times 5\times 9$ mm${^3}$, respectively.
Under the C$_1$ symmetry and through the hyperfine interactions, the level degeneracy of $^{171}$Yb ions ($S=1/2$, $I=1/2$) in both sites is completely lifted and  each crystal field (CF) level splits into 4 hyperfine sublevels or spin states\,\cite{Ortu2018}.  The partially-resolved 16 optical hyperfine transitions between the $^2\mathrm{F}_{7/2}(0)$ and $^2\mathrm{F}_{5/2}(0)$ CF levels of site II are displayed in the absorption spectrum in Fig.\,\ref{fig:absorption}. 
We chose to focus our study on $^{171}$Yb ions in site II, which has been the focus of recent work on quantum memories due to its higher absorption strength\,\cite{Welinski2016}. In particular, we probed the coherence properties of the transition at $\nu$=306263.5\,GHz (978.85 nm in vacuum) between the highest hyperfine level of the ground state (4g) and the lowest of the excited level (1e) since it is the most spectrally isolated. The chances of simultaneously exciting several transitions with potentially different properties are thus minimized. The light propagated along the b axis and was polarized along the crystals $D_2$ dielectric axis to maximize the absorption. 

\section{Decoherence processes of optical transitions}

The optical homogeneous linewidth is determined by the ground $\mathrm{T_{1,4g}}$ and excited $\mathrm{T_{1,1e}}$ level population lifetimes and by the pure dephasing rate $\Gamma_\phi$ as follows:
\begin{equation}
\Gamma_h = \frac{1}{2 \pi \mathrm{T_{1,4g}}} + \frac{1}{2 \pi \mathrm{T_{1,1e}}} + \Gamma_{\phi}.
\label{Eq:linewidth_broadening0}
\end{equation}

Note that the linewidths are here expressed as FMHW in units of Hz. The first two terms are referred to as \emph{direct} or $T_1$-type  contributions to $\Gamma_h$. 
The lifetime of a ground state hyperfine level $4g$ is defined as follows:
\begin{equation}
    \frac{1}{\mathrm{T_{1,4g}}} = \sum_{j\neq 4}R_{s,4g \rightarrow jg} + 
    R_{\textrm{ff},(4g, jg) \rightarrow (jg,4g)}.
\end{equation}
The first and second right-hand side terms represent spin-lattice relaxation (SLR) and flip-flop (FF) processes, summed over all other ground state hyperfine levels. The lifetimes of the excited state hyperfine levels are expressed in the same way, with the addition of the optical relaxation rate $\mathrm{1/T_\textrm{opt}}$. The $\mathrm{1/T_\text{opt}}$ rate is essentially radiative as the multiphonon relaxation from the excited $^{2}\mathrm{F}_{5/2}(0)$ level to the ground $^{2}\mathrm{F}_{7/2}$ multiplet is strongly hampered by the large energy gap ($\sim10000\,$cm$^{-1}$)\,\cite{miyakawa1970phonon,thibault2008growth}.The relaxation is characterized by branching ratios $\beta_{ie,jg}$ specific to the involved hyperfine levels \cite{Tiranov2018}. 

SLR itself includes several mechanisms. In the one-phonon direct process, the spin-flip is accompanied by the emission or absorption of one phonon, and its rate is strongly dependent on the phonon energy $\hbar \omega$ required to bridge the energy gap between the spin levels. In our system and conditions, i.e.  Kramers conjugate spin levels and zero magnetic field, it can happen through state admixture by the hyperfine interaction \cite{bakerDependenceSpinLatticeRelaxation1964}.  In the two-phonon Raman scattering, the spin-flip is induced by a virtual absorption of one phonon followed by the virtual emission of another phonon. For the Kramer ions, this process scales with the ninth power of the temperature $T$, for $T\ll T_D$, with $T_D$ the crystal Debye temperature. Finally, if there exists a phonon mode resonant to a CF transition of energy $\Delta$, an Orbach process takes place in which phonon absorption-emission is not virtual and spin-phonon coupling strength is greatly enhanced. 
In the limit $kT \gg \Delta_{i,j}$, where $k$ is the Boltzmann constant and  $\Delta_{i,j}$ the energy difference between the $i$ and $j$ hyperfine levels, the overall SLR rate can be thus expressed as\,\cite{abragam2012electron}:
\begin{equation}
\label{Eq:SLR}
R_{\textrm{s},i \rightarrow j} = \alpha_{d,i,j} T + \alpha_{R,i,j} T^9 + \frac{\alpha_{O,i,j}} {\exp(\Delta/kT)-1}. 
\end{equation}
Here, $T$ is the temperature, $\hbar$ the reduced Planck constant, and $\alpha_{d,i,j}$, $\alpha_{R,i,j}$, and $\alpha_{O,i,j}$ are coefficients accounting for the strength of each process. 

Besides the spin-phonon interactions, the spin lifetime can also be shortened by state exchange between resonant spins via the flip-flop process induced by dipole-dipole interactions. The rate for an ion in level $i$ to exchange state with a nearby ion in level $j$ ($(i,j) \rightarrow (j,i) $) depends on the concentration of the ions, the wavefunction of the two levels and the  the spin inhomogeneous transition linewidth\,\cite{car2019optical,Bottger2006,abragam2012electron,Welinski2020}.
In section \ref{sec:spin_lifetime}, hyperfine level populations dynamics are modeled for SLR and FF processes with a number of simplifying assumptions such as $\alpha_{R,i,j} = \alpha_R/3$, i.e. no dependence of the Raman process on hyperfine levels.

The pure dephasing rate $\Gamma_\phi$ can be written as:
\begin{equation}
\Gamma_{\phi} =  \Gamma_\textrm{Yb-Ph} + \Gamma_\textrm{Yb-Imp} +  \Gamma_\textrm{Yb-Yb} + \Gamma_\textrm{Yb-Y} +\Gamma_\textrm{others}.
\label{Eq:linewidth_broadening}
\end{equation}
The equation includes  $^{171}$Yb interactions with phonons $\Gamma_\textrm{Yb-Ph}$, with impurities or defects $\Gamma_\textrm{Yb-Imp}$, with electron spin of other Yb ions $\Gamma_\textrm{Yb-Yb}$ and with nuclear spins $\Gamma_\textrm{Yb-Y}$, that in YSO are predominantly represented by $^{89}$Y$^{3+}$ ($I=1/2$, 100\% abundance) ions. Interaction with phonons may perturb the coherent ions phase through the elastic two-phonon processes like the Raman process giving a $T^7$ temperature dependence at low temperatures for a Debye phonon density of states\,\cite{konzTemperatureConcentrationDependence2003,Bottger2016,mccumberLinewidthTemperatureShift1963}. Interactions with local vibration modes have also been reported, leading to a temperature dependence similar to the Orbach process, where $\Delta$ is the local mode energy \cite{chiossi2022photon}. The next two terms $\Gamma_\textrm{Yb-Imp}$ and $\Gamma_\textrm{Yb-Yb}$ in Eq. \ref{Eq:linewidth_broadening} account for the magnetic noise experienced by the coherently probed ions driven by the relaxation of impurities/defects and nearby Yb ions. These variations in the coherent ions environment lead to a spreading of the frequency transitions, a phenomenon known as spectral diffusion. $^{171}$Yb spins in sites I and II can participate in this magnetic noise so that the SLR and FF processes described above also have an \emph{indirect} or $T_2$-type contribution through the pure dephasing term $\Gamma_h$.

Among the other broadening factors $\Gamma_\textrm{others}$ in Eq.\,\ref{Eq:linewidth_broadening}, one should consider the instantaneous spectral diffusion and effects of the two-level systems (TLS) which have been encountered in some RE doped crystals and ceramics\,\cite{graf1998photon,gritsch2022narrow,Huber1984,fukumori2020subkilohertz,kunkel2016dephasing,flinnSampledependentOpticalDephasing1994,macfarlaneOpticalDephasingDisorder2004}. The first effect considers the frequency shift of the excited ions due to the laser excitation of other nearby Yb ions. It is independent of the temperature and is generally suppressed in very low-concentrated crystals. The TLS mode, namely the tunneling between two nearly-degenerate states or atomic configurations, may also decrease the coherence time. For this mechanism, a characteristic $\sim T^{1.3}$ dependence was calculated and observed\,\cite{Huber1984,Macfarlane2000,macfarlaneOpticalDephasingDisorder2004}.

\section{Results}
The understanding of $^{171}$Yb optical dephasing requires knowledge of the dynamics of $^{171}$Yb spins. Spin relaxation reduces the lifetime of the optical and ground state of the coherent ions, thereby broadening the homogeneous linewidth via a direct mechanism. The first subsection\,\ref{sec:spin_lifetime} is thus devoted to estimating spin lifetime between 40\,mK and 10\,K for both $^{171}$Yb:YSO samples. We employed the SHB and the 3PE techniques and distinguished two regimes, i.e. below 5\,K and above 6\,K in which the spin relaxation is dominated by the FF processes and SLR, respectively. We focused on relaxations occurring on a few ms timescale as this is the maximum timescale over which homogeneous linewidths may be affected.

In subsection\,\ref{sec:coherence_time}, we exploited two-pulses photon echo (2PE) and SHB to measure the optical homogeneous linewidth of the $^{171}$Yb ions in the two samples between 40\,mK and 18\,K. The results allow us to discriminate and quantify the different types of interaction causing the optical decoherence on the basis of the temperature and doping concentration dependencies. The line broadening at the long timescale due to spectral diffusion at different temperatures was investigated using 3PE.

\subsection{Spin population dynamics}
\label{sec:spin_lifetime}
Spin population relaxation was determined by burning a spectral hole in the 4g-1e optical transition and monitoring the evolution of hole area, which is insensitive to spectral diffusion, after a varying delay time $\mathrm{T_D}$. The hole area is proportional to the deviation of the 4g level population from its thermal equilibrium value: $n_{4g}-n_{4g}^{eq}$. It corresponds to the optically pumped ions that have not returned to their initial level. Its evolution thus depends on the decay of the excited state hyperfine levels to the ground ones through optical relaxation and the relaxations within the ground state hyperfine levels. Since the optical transitions between the excited and ground state hyperfine levels are not of equal strength \cite{Tiranov2018}, relaxation between the excited state hyperfine levels can also play a role in the spectral hole area decay.  It therefore appears that many parameters are needed to fully describe these processes that moreover may include FF and SLR, depending on the temperatures and Yb concentrations considered. 

In order to reduce the level of complexity, we took advantage of the strong temperature dependence of SLR described above.  Relaxation processes of the hole area decay were modeled on the one hand for "low" temperatures. Here, only FF is considered in addition to optical relaxation, and SLR is neglected. On the other hand, in the "high" temperature regime, only SLR in the ground and excited spin states are taken into account, again in addition to optical relaxation. These models are described in detail in SM sec.\,\ref{sec:model_sf}.

\begin{figure}[h!]
	\includegraphics[width=1\linewidth]{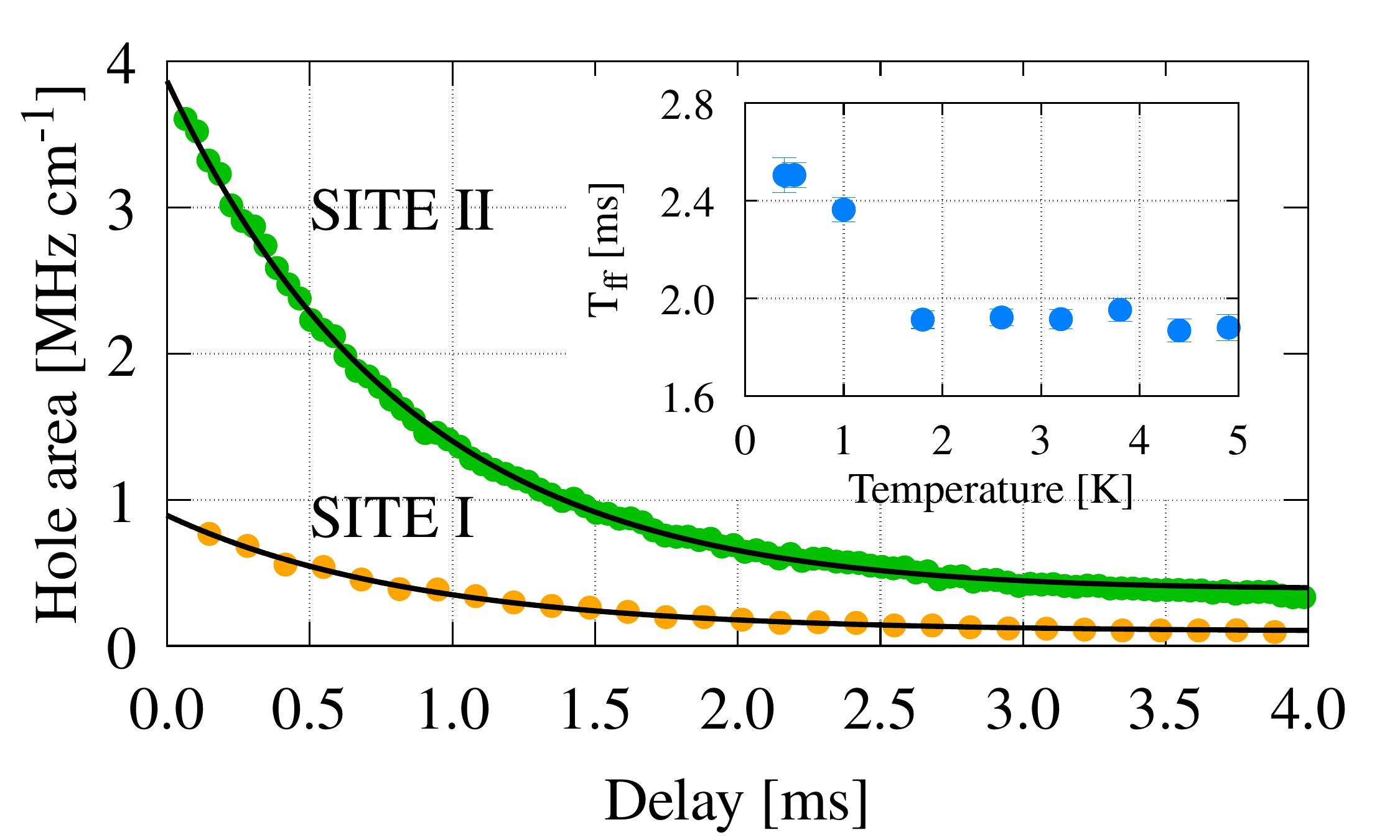}
	\caption{Time dependence of the spectral hole area obtained after burning in the transitions 4g$\rightarrow$1e of site II at 306263.7\,GHz (green circles) and 1g$\rightarrow$4e of  site I at 305459.4\,GHz (orange circles) in the 10 ppm sample and at 1.8 K. Black lines are fits using the "low" temperature model where FF dominate relaxation (see SM sec.\,\ref{sec:model2}). Inset: fitted spin relaxations times of Yb ions in site II as a function of the base plate temperature.}
	\label{fig:ff}
\end{figure}  

\subsubsection{Low temperature regime}
\label{sec:spinLT}
We first investigated temperatures below 5 K in the 10 ppm sample with the low-temperature model.
Due to the high anisotropy of the YSO matrix, very different rates are expected for the FF processes depending  on the initial states of the involved Yb ions. In particular, a rate of 390\,s$^{-1}$ was theoretically estimated for the $1g\rightleftarrows2g$ and $3g\rightleftarrows4g$ processes in the 10\,ppm sample while other state combinations give  rates at least two orders of magnitude slower\,\cite{Welinski2020}.
No FF should occur between ions in the optically excited states since only a small fraction (less than 0.2\%) of the ions are pumped in the 1e level while FF requires significant population density and at least two populated levels. 

Our model of the 4g-1e hole area decay at delays of a few ms, which is relevant to the optical decoherence processes studied here, thus takes into account only the fast $3g\rightleftarrows4g$ process and the optical state relaxation. Under these assumptions, the hole area $H$ is proportional to the 4g level population deviation from equilibrium given by:
\begin{eqnarray}
H(t) &=& c(n_\mathrm{4g}^\mathrm{eq}-n_\mathrm{4g}(t)) \nonumber \\  &=& cN\{e^{-2t/\mathrm{T_{ff}}}+A(e^{-t/\mathrm{T_{opt}}}-e^{-2t/\mathrm{T_{ff}}}) \notag \\ 
&& +B(1-e^{-2t/\mathrm{T_{ff}}})\},
\label{Eq:20} 
\end{eqnarray}
where N is the initial 1e level population, and {$\mathrm{T_\text{ff}}=1/R_{\mathrm{ff},(4g,3g)\rightarrow (3g,4g)}$} is the inverse of the 4g$\rightleftarrows$3g FF rate for the ions in level 4g. The two last terms on the right-hand side are due to relaxation from the excited state. The $A$ and $B$ constants are given in SM sec.\,\ref{sec:model1} and depend on $\mathrm{T_\text{opt}}$, $\mathrm{T_\text{ff}}$, and the branching ratios between the excited and ground state hyperfine levels\,\cite{Nicolas2022coherent}. 
Only two parameters were varied to model experimental data using Eq.\ref{Eq:20}, i.e. $\mathrm{T_{ff}}$ and a scaling factor $cN$ corresponding to $H(0)$. $\mathrm{T_\mathrm{opt}}$ was fixed at the measured value of 1.3 ms \cite{Welinski2016} and the branching ratios were taken from the experiments of Ref. \onlinecite{Tiranov2018,Nicolas2022coherent}. A good fit to the data collected at 1.8\,K for the 10\,ppm sample is obtained, as shown in Fig.\,\ref{fig:ff}, and gives $\mathrm{T_{ff}}=1.91\,\pm0.04$\,ms. This value is in fair agreement with the calculated value (1/390\,s$^{-1}$=2.5\,ms) and the experimental one ($\sim2\,$ms) obtained in Ref.\,\onlinecite{Welinski2020}. 

Hole decays recorded up to 5\,K  could be well fitted with our model. It resulted, within experimental uncertainty, in a temperature-independent $\mathrm{T_\text{ff}}$. Below 1\,K, the thermal polarization of spins in lower energy hyperfine ground states decreases FF process involving the 4g and 3g levels. Indeed,  $\mathrm{T_\text{ff}}$ increases with decreasing temperature (Fig. \ref{fig:ff}, inset) but not to the expected extent. By analyzing the absorption spectra at base temperatures at 40\,mK, we found out that the spin level populations do not match the expected Boltzmann distributions but correspond to higher sample temperatures up to several hundreds of mK. This could be related to poor thermalization of the spin in the crystal at ultra-low temperatures \cite{Kukharchyk2018}.

Spectral hole decay measurements were also performed for Yb ions in site I by burning the transition 1g-4e at 305457.4\,GHz. As shown in Fig.\,\ref{fig:ff} in orange, the hole area decay is well fitted up to 4\,ms with the formula $A(t)=a+b\cdot\exp(-t/\mathrm{T_{opt}})$ where $\mathrm{T_{opt}}$ for site I ions is fixed at 0.87\,ms\,\cite{Welinski2016}. This implies that the hole area decay at short times is mainly due to the optical relaxation, while FF has a much longer characteristic time. Indeed, by recording the spectral hole for longer delays, we estimated a component in the hole area decay of tens of milliseconds and a longer component that makes the hole live for more than 1\,s. The FF rate of Yb ions in site I is therefore much slower compared to the rate of ions in site II. 

In the 2\,ppm sample, the hole area decay measured for site II ions is completely determined by the optical state relaxation in the first ten milliseconds up to 5\,K and the hole relaxation due to FF processes has a much longer characteristic time. 


\subsubsection{High temperature regime}
\label{SpinHT}
For temperatures above 6 K, a strong dependence of spectral hole decay on temperature was observed in the 2 and 10\,ppm samples, indicating the predominance of SLR. Under these conditions, we fitted, at varying temperatures, the hole area decays  to a model that includes SLR for ground and excited hyperfine levels, and optical relaxation. To further simplify expressions and reduce the number of free parameters, SLR rates are assumed identical for all ground and excited hyperfine levels. {Let $1/3\mathrm{T_S} = R_{s,4g,jg}=R_{s,1e,je}$ be the SLR relaxation rate from the 4g or 1e level to any other $jg$ or $je$ levels, respectively. Then, $1/\mathrm{T_S}$ is the overall SLR rate for both 1e and 4g levels.}

The population of level 4g then reads (see SM sec.\,\ref{sec:model2}):
\begin{align}
n_\mathrm{4g}^\mathrm{eq}-n_\mathrm{4g}(t) &= N(e^{-4t/3\mathrm{T_S}}+ A'e^{-4t/3\mathrm{T_S}}[1-e^{-t/\mathrm{T_{opt}}} ]  \nonumber \\ &+
B'[e^{-t/\mathrm{T_{opt}}}-e^{-4t/3\mathrm{T_S}}])
\label{Eq:population4g},
\end{align}
As in the low-temperature model, the $A'$ and $B'$ constants depend on $\mathrm{T_\text{opt}}$, $\mathrm{T_S}$, and the branching ratios between the excited and ground state hyperfine levels. Eq.\,\ref{Eq:20} and Eq.\,\ref{Eq:population4g} also indicate that spectral hole decays are not single exponential in general. 

\begin{figure}[h!]
	\includegraphics[width=1\linewidth]{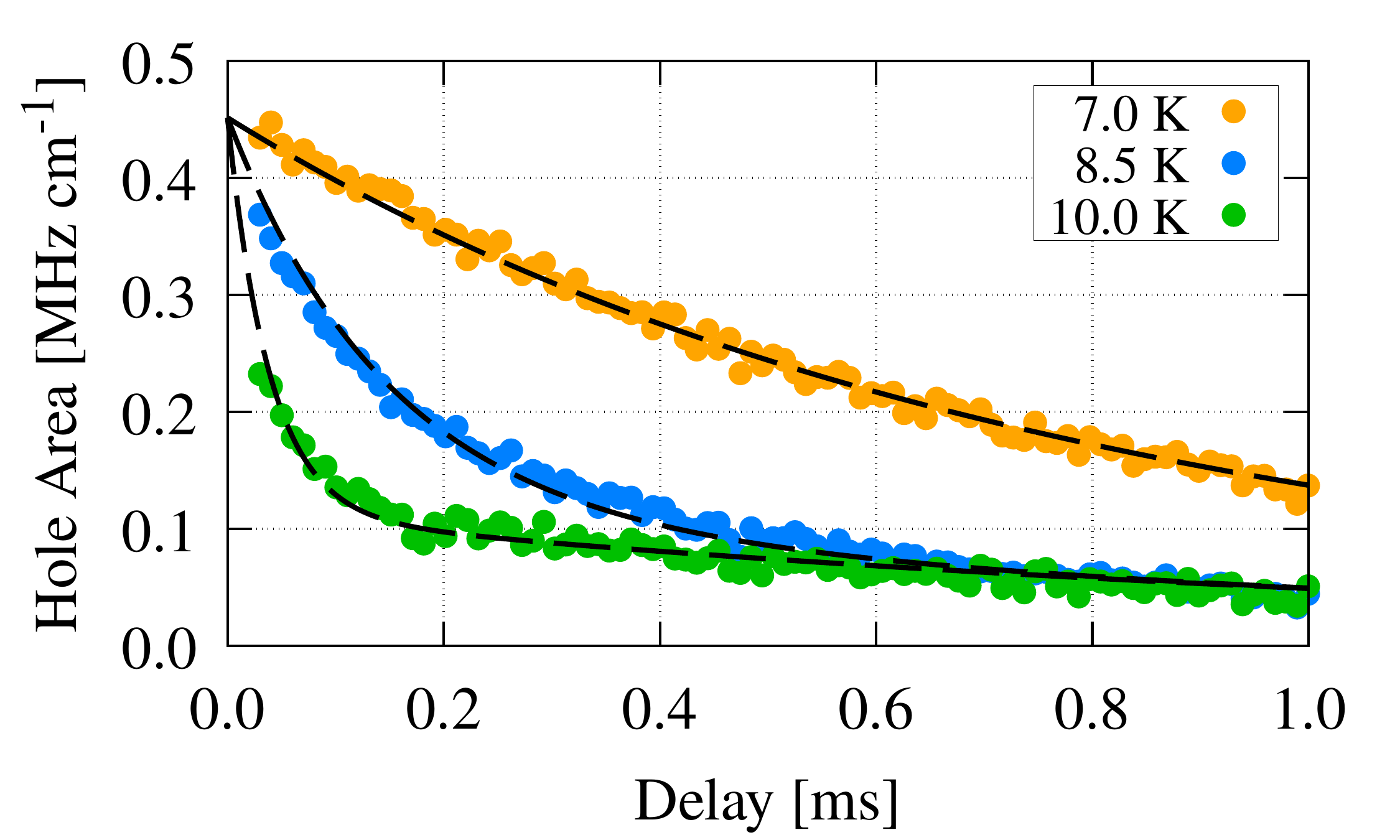}
	\caption{Spectral hole area decays in the $^{171}$Yb:YSO\,2\,ppm sample for three temperatures. The dashed lines are the least-squares fits to the data using eq\,\ref{Eq:population4g}.}
	\label{fig:SHB_decay}
\end{figure}

\begin{figure}[h!]
	\includegraphics[width=1\linewidth]{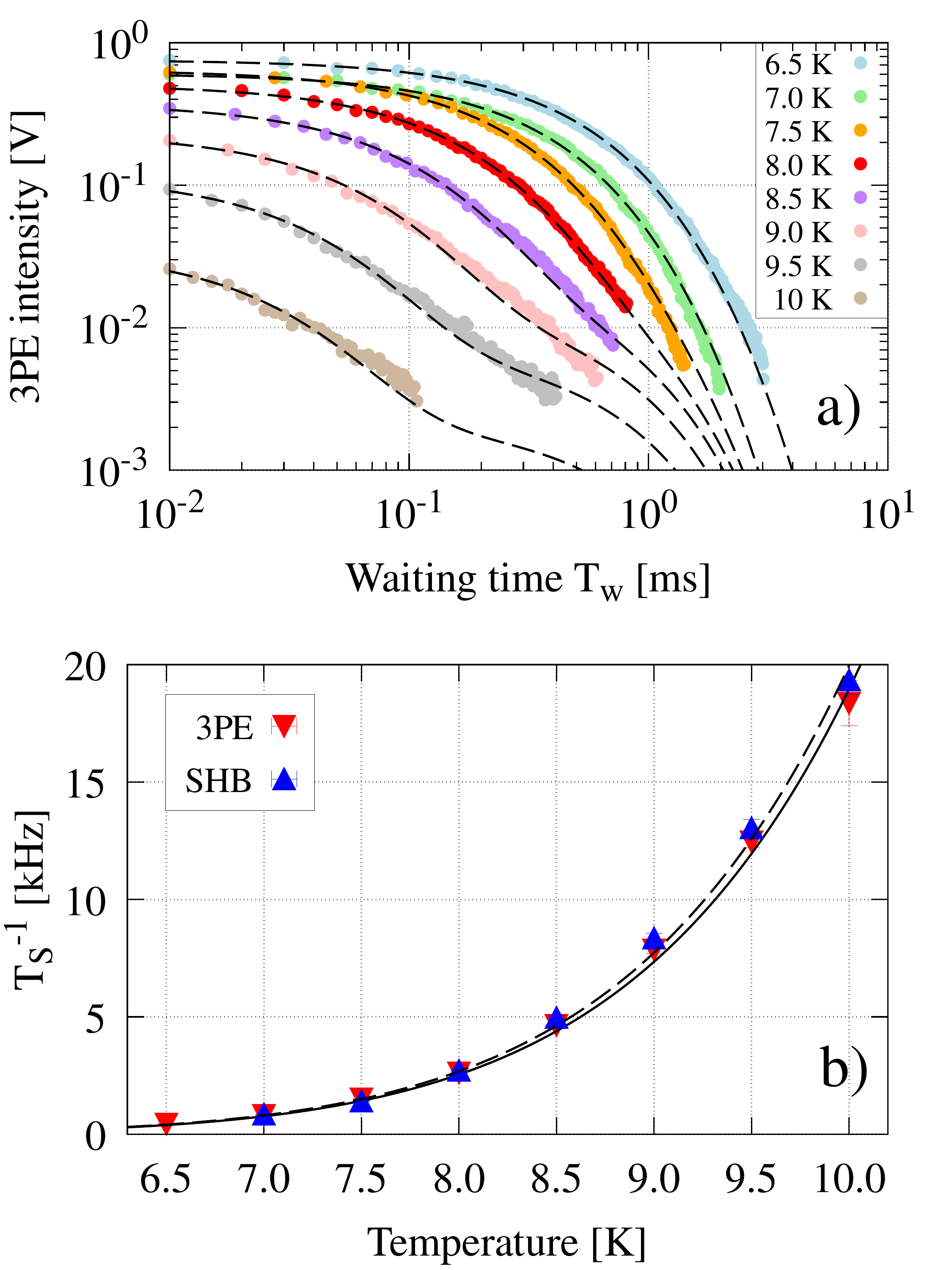}
	\caption{a) 3PE intensity in the $^{171}$Yb:YSO 2\,ppm sample for $\tau_d=7\,\mu$s, varying the waiting time $\mathrm{T_W}$ and the crystal temperature. The dashed black lines represent fits to the data using Eq.\,\ref{Eq:suppl_3PE_model} in SM. b) Estimated spin-flip rates obtained by the analysis of the 3PE decay (red) and spectral hole area decay (blue) as a function of the temperature. The fit to the red points is displayed with a continuous black line while the one to the blue points with a dashed line.}
	\label{fig:3PE_decay}
\end{figure}

SLR at high temperatures does not depend on Yb concentration so we fitted the hole area decay recorded in the 2\,ppm sample, as the FF rate is already negligible at 5\,K in this case. Since we used identical and short burning pulses in all measurements, the initial hole area was the same for all data sets and estimated to be 0.45\,MHz\,cm$^{-1}$ by fitting the data at 6\,K.  It was then fixed for all the other data sets, leaving $\mathrm{T_{S}}$ as the only free parameter. It is noticeable, as shown for a few temperatures between 6.5 and 10\,K in Fig.\,\ref{fig:SHB_decay}, that the data show only one characteristic SLR component for any temperature and are well-fitted by our simplified model, suggesting that different pairs of states have indeed similar SLR rate. 
The obtained $\mathrm{T_{S}}$ values range between 3\,ms and 50\,$\mu$s. The SLR rate values $1/\mathrm{T_{S}}$ of the 4g and 1e levels are plotted in Fig.\,\ref{fig:3PE_decay}b, as a function of temperature and were found to follow the function $\alpha_RT^9$ with $\alpha_R=(2.00\pm0.03)\cdot10^{-5}$\,Hz/K$^9$.

This temperature dependence clearly indicates that the phonon Raman scattering term in Eq.\,\ref{Eq:SLR} is much larger than the other terms, at least for temperatures higher than 6\,K. Actually, both the ground state and excited level of the probed transition are separated by more than 230\,cm$^{-1}$ from the next Stark level. The Orbach process is therefore strongly suppressed below 20\,K for which the thermal energy $k_B T$ is only 14 cm$^{-1}$. The use of a single $1/3\mathrm{T_S}$ rate for all hyperfine level pairs in our model could be then partially justified by the fact that Raman scattering is independent of the energy splitting between the involved states\,\cite{abragam2012electron}. 


The SLR rate can also be measured by analyzing three-pulse photon echo (3PE) experiments which use the pulse sequence ($\pi/2-\tau_d-\pi/2-\mathrm{T_W}-\pi/2-\tau_d-\mathrm{echo}$), with $\mathrm{T_W}$ the waiting time between the second and third laser pulse. Compared to hole burning that often requires significant pumping time (25 $\mu$s in the above experiments), this method can probe population dynamics on the time scale of the $\pi$/2 pulses (1 $\mu$s in our case). The 3PE intensity is given by \cite{ahlefeldt2015}:


\begin{equation}
	\mathrm{I}(\tau_d,\mathrm{T_W}) = \frac{\mathrm{I_0}}{4} (n_\textrm{1e} + n^{eq}_\textrm{4g}-n_\textrm{4g})^2 \mathrm{exp}(-4\pi \tau_d \Gamma_\mathrm{eff}),
	\label{Eq:3PE_2} 
\end{equation}
where $n_{1e}$ is the 1e level population. It is given by:
\begin{eqnarray}
   {n_\mathrm{1e}}(t) = \dfrac{N}{4}e^{-t/{\mathrm{T_{opt}}}} [ 3e^{-4t/\mathrm{3T_{S}}}+1  ]
\label{Eq:boh1}
\end{eqnarray}
under the assumptions leading to Eq. \ref{Eq:population4g}.
The effective linewidth $\Gamma_\mathrm{eff}=\Gamma_h+\Gamma_\mathrm{SD}(\tau_d,\mathrm{T_W})$ takes into account the additional homogeneous broadening, known as spectral diffusion, occurring during the waiting time $\mathrm{T_{W}}$.

As shown later in sec.\,\ref{sec:spectral_diffusion}, the spectral diffusion for the 2\,ppm sample is at most a few kHz. For $\tau_d\leq 10$ $\mu$s,   the exponential term in Eq.\,\ref{Eq:3PE_2} can therefore be considered independent of $\mathrm{T_W}$.  
At fixed $\tau_d$,  Eq. \,\ref{Eq:3PE_2} then becomes:
\begin{equation}
	\mathrm{I}(\mathrm{T_W}) = \frac{1}{4}\mathrm{I_0'} (n_\textrm{1e} + n^{eq}_\textrm{4g}-n_\textrm{4g})^2
	\label{eq:3PE_semp}
\end{equation}
in which the squared term is deduced from Eq. \ref{Eq:population4g} and \ref{Eq:boh1} (see SM sec.\,\ref{sec:model_sf}).

We recorded the 3PE peak intensity fixing $\tau_d=7\,\mu$s, for different temperatures and waiting times. The data for each temperature above 6\,K was fitted with Eq.\,\ref{eq:3PE_semp} and are shown in Fig.\,\ref{fig:3PE_decay}a. Note that even if $\mathrm{T_{S}}$ is the only free parameter of the model besides a scaling factor, data are well fitted, confirming again the validity of the assumptions employed. The obtained $1/\mathrm{T_{S}}$ values are plotted in Fig.\,\ref{fig:3PE_decay}b as a function of temperature and fitted using the function $\alpha_RT^9$, obtaining $\alpha_R=(1.90\,\pm0.03)\cdot10^{-5}$\,Hz/K$^9$. We notice that the lifetimes and Raman process coefficients derived from 3PE and SHB measurements are very similar. Although driven by the same mechanisms,  hole area and 3PE intensities have quite different dependencies on the ground and excited populations and the excellent agreement of the results highlights the suitability of the proposed model.
		
\subsection{Optical coherence}
\label{sec:coherence_time}

	\begin{figure}[h!]
	\includegraphics[width=1\linewidth]{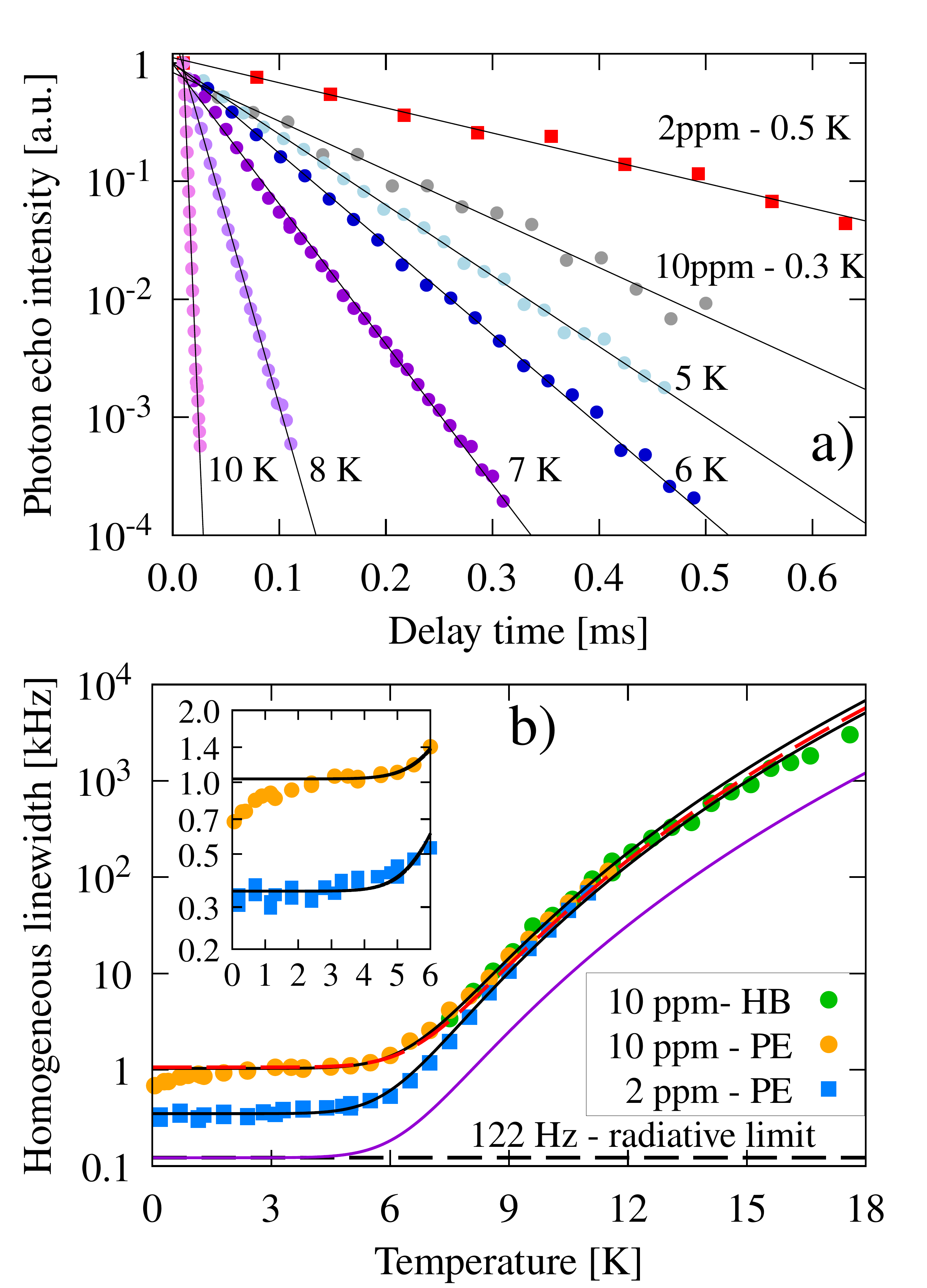}
	\caption{2PE experiments. a) Normalized photon echo peak intensity as a function of the delay time for the Yb:YSO 2\,ppm (squares) and 10\,ppm (circles) samples at different temperatures. The black lines represent exponential fits. b) Homogeneous linewidths measured for the two samples at different temperatures using the spectral hole burning (squares) and the 2PE (circles) techniques. The continuous black lines and red dashed lines are fits to the 2PE and SHB data. The violet line represents the dephasing contribution due to the $T_1$ mechanism for the 2\,ppm sample calculated as $(2\pi \mathrm{T}_\mathrm{opt})^{-1}+\alpha_RT^9/\pi$.}
	\label{fig:Coherence_temperature}
\end{figure}

\subsubsection{Two-pulse echo measurements and spectral hole burning}
\label{sec:2PE}

The details of the 2PE pulse sequence $\pi/2-\tau_d-\pi-\tau_d-\text{echo}$ and the measures taken to mitigate or compensate for the vibration-induced signal fluctuations, as well as the hole-burning effect, are described in Appendix. The obtained 2PE echo peak intensities at different temperatures $T$ are plotted in Fig.\,\ref{fig:Coherence_temperature}a for the 2 and 10 ppm samples. The 2PE decay is purely exponential and the homogeneous linewidth $\Gamma_h$ can be estimated using the formula: $I(\tau_d) = I_0\textrm{exp}(-4\tau_d \pi \Gamma_h)$,  where $\Gamma_h$ is the FWHM homogeneous linewidth in units of Hz. The fitted homogeneous linewidths for the two crystals s a function of $T$ are shown in Fig.\,\ref{fig:Coherence_temperature}b. 

The Yb:YSO 2\,ppm sample exhibits an exceptionally narrow linewidth $\Gamma_h=(320\pm20)\,$Hz at ultra-cryogenic temperatures, corresponding to a coherence time of around 1\,ms - one of the longest measured for RE ions. The homogeneous linewidth is nearly constant up to 3\,K, slightly increases between 3 and 6\,K, and grows rapidly only above 6\,K. A similar behavior is found for the 10 ppm Yb:YSO sample but the homogeneous linewidth in the (2-5)\,K is $\sim$1\,kHz and it decreases for lower temperatures reaching 700\,Hz at 40\,mK. 

Between 6 and 11\,K, $\Gamma_h$ in the two crystals has very similar trends even if the values for the 10\,ppm remain slightly broader than the ones of the 2\,ppm sample. It can be noticed that narrow homogeneous linewidths are retained even at elevated temperatures. For example, a 25\,kHz linewidth is measured at about 10\,K. This could be of interest for radio-frequency signal analysis based on SHB, which requires this kind of resolution for communication signals. As the Yb optical transitions are within the therapeutic window, the possibility of burning narrow holes at high temperatures could be also useful for medical ultra-sound optical tomography\,\cite{thai2022vivo}. 

As displayed in Fig.\,\ref{fig:Coherence_temperature}, the homogeneous linewidth  of the 2\,ppm sample and the one of the 10\,ppm sample above 2\,K can be well fitted by the function:  $\Gamma_h(T) = \Gamma_h(0)+\gamma_RT^9$. The fit results for both samples are reported in table\,\ref{tab:fit_results}. 

For temperatures above 11\,K coherence times became too short to be measured with the 2PE method. To extend our investigation to higher temperatures, we employed the SHB technique. We selectively burnt the narrowest  possible hole in the 4g-1e transition of the 10\,ppm sample by using short and weak burning pulses, and short delays between burning and read-out pulses. This reduced influence of laser contribution $\Gamma_\text{laser}$, spectral diffusion, and power broadening on the measured hole width $\Gamma_\text{hole}$ (see Appendix). Using the formula $\Gamma_h = (\Gamma_\text{hole}-2\Gamma_\text{laser})/2$, where $\Gamma_\text{laser} = 28$ kHz, homogeneous linewidths could be reliably determined between 7 and 18 \,K (Fig.\,\ref{fig:Coherence_temperature}). The two techniques lead to the same linewidth values in the common temperature range (7-11\,K), supporting the choice of the SHB experimental parameters. A good fit of the SHB results was obtained using the formula $\Gamma_h(T) = \Gamma_h(0)+\gamma_RT^9$, in which $\Gamma_h(0)$ was set at 1.03\,kHz. The fitted $\gamma_R$ coefficient is comparable to the  values estimated by 2PE (see Table.\,\ref{tab:fit_results}).



\begin{table}[ht!]
	\centering
		\caption{Best fit parameters for the temperature dependence of homogeneous linewidth measured between 40\,mK and 18\,K.}
	\label{tab:fit_results}
	\small
		\begin{threeparttable}
			
			\begin{tabular}{@{} *1l*3c @{}}   
				
				\textbf{} & Method  & \,\,\, ${\Gamma_h(0)}$ [kHz] \,\,\,& {$\gamma_R$} [10$^{-5}$\,Hz/K$^{9}$]   \\ \midrule		
				
				\textbf{Site II 2\,ppm } & 2PE & 0.35$\pm$0.01 & 2.57$\pm$0.08 \\
				\textbf{Site II 10\,ppm } & 2PE & 1.03$\pm$0.02 & 3.45$\pm$0.03  \\
				\textbf{Site II 10\,ppm } &  SHB & - & 2.88$\pm$0.15  \\

			\end{tabular}
		\end{threeparttable}
\end{table}


\subsubsection{Three-pulse echo measurements}
\label{sec:spectral_diffusion}



\begin{figure}[h!]
	\includegraphics[width=1\linewidth]{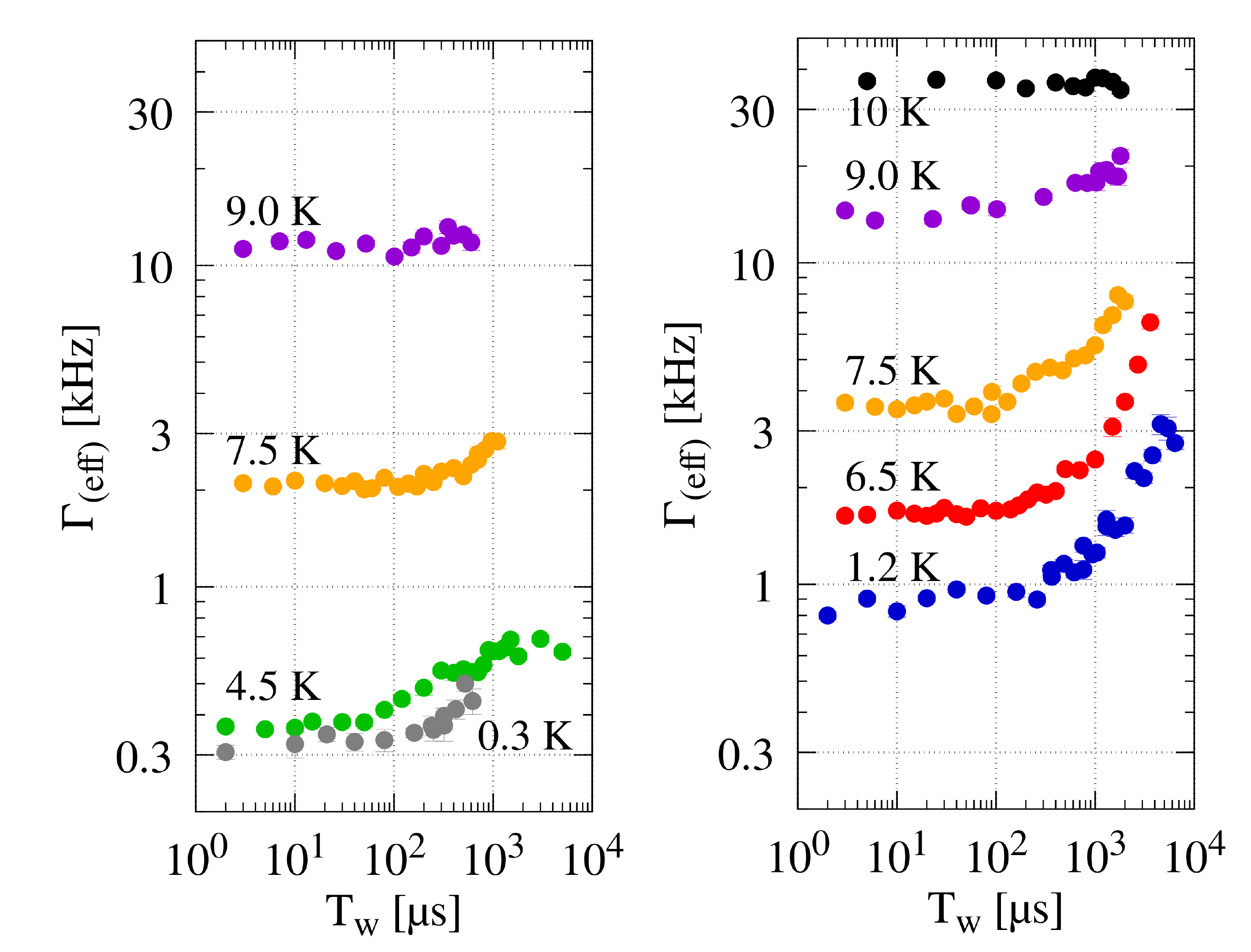}
	\caption{Evolution of the effective linewidth in $^{171}$Yb:YSO 2\,ppm (left) and 10\,ppm (right) samples for the 4g$\rightarrow$1e transition and temperatures between 300\,mK and 10\,K.}
	\label{fig:spectral_diffusion}
\end{figure}

We finally investigated long-timescale evolutions of homogeneous linewidths using  3PE experiments. Unlike the measurements described in Sec. \ref{SpinHT}, we measured the 3PE intensity for fixed $\mathrm{T_W}$ values and varying $\tau_d$. The decays were found to be well fitted by single exponentials, indicating that the effective homogeneous linewidth $\Gamma_\text{eff}$ in Eq. \ref{fig:3PE_decay} only depends on $\mathrm{T_W}$.  
$\Gamma_\mathrm{eff}$ values as a function of $\mathrm{T_W}$ for the two samples and different temperatures are shown in Fig.\,\ref{fig:spectral_diffusion}. 
By inspecting Fig.\,\ref{fig:spectral_diffusion}, it can be seen that $\Gamma_{\mathrm{eff}}$ approaches $\Gamma_h$ at the shortest $\mathrm{T_W}$ in both samples, as expected, but broadens appreciably for increasing $\mathrm{T_W}$. 

In the 2\,ppm sample, broadening sets in on a 100s of $\mu$s time scale and varies from 200 Hz at 300 mK to 1 kHz at 7.5\,K. At 9K, a constant linewidth of about 10 kHz  is observed as a function of $\mathrm{T_W}$. The 10 ppm crystal exhibits a similar behavior but with a larger broadening of $\Gamma_\text{eff}$ that reaches 2 kHz at 1.2 K up to 7\,kHz at 9 K. The effective linewidth at 10 K is 37 kHz and does not broaden as a function of $\mathrm{T_W}$.

Spectral diffusion could also be evaluated by monitoring the broadening of narrow spectral holes. 
Unfortunately, the laser limitation on the hole linewidth prevented us to determine any small transient and persistent hole broadening. Nonetheless, our SHB experiments excludes spectral diffusion larger than a few tens of kHz for the 10\,ppm sample, confirming the constraint set by the 3PE experiment. 



\section{Discussion}
In this section, we discuss the decoherence mechanisms affecting $^{171}$Yb optical transition in the 2 and 10 ppm samples in different temperature regimes. 

\subsection{Direct contribution to decoherence}

The comparison between $^{171}$Yb spin population relaxation rates and optical homogeneous linewidths shows very similar trends:  1) between 2 and 5\,K, both of them have constant values, with larger values found in the 10 ppm sample; 2) above 6\,K a fast increase is observed with close temperature dependence; 3) below 2\,K, a small decrease in relaxation rate and linewidth is observed in the 10 ppm sample. This clearly suggests that decoherence is driven to a large extent by $^{171}$Yb spin relaxation, except possibly for the lower doping level and temperatures (see Sec.\,\ref{sec:subsectionB}). 

The spin relaxations measurements and modeling of Sec. \ref{sec:spin_lifetime}, allow calculating the direct contribution to $\Gamma_h$ through the population decays of the 4g and 1e levels. Using Eqs. \ref{Eq:20}, \ref{Eq:population4g}, and \ref{Eq:boh1}, we estimated it in the following way: 
\begin{eqnarray}
    \Gamma_\text{direct, LT} &=& \frac{1}{2\pi \mathrm{T_\text{opt}}}+ \frac{1}{2\pi \mathrm{T_\text{ff}}}\quad \text {and}\\
    \Gamma_\text{direct, HT} &=& \frac{1}{2\pi \mathrm{T_\text{opt}}}+ \frac{2}{2\pi \mathrm{T_\text{S}}}
\end{eqnarray}
for the low and high temperatures regimes.
Fig. \ref{fig:Coherence_temperature} shows the dependence of $\Gamma_\text{direct}$ on temperature. It is clear that it cannot fully account for the observed homogeneous linewidths in the whole studied temperature range. For example, the 320\,Hz measured at the lowest temperatures in the 2\,ppm sample exceeds by 200\,Hz the radiative limit $\Gamma_\mathrm{opt}=122$\,Hz. 
In the high temperature regime, the direct contribution from the $T^9$ Raman process accounts for only 24\% of the observed total linewidth.

\subsection{Low temperature regime ($< 5$ K)}
\label{sec:subsectionB}
The optical homogeneous linewidth of the Yb ions in the 2\,ppm sample does not change below 2\,K, where spin polarization is expected, in contrast with the 10 ppm sample behavior (see Fig. \ref{fig:Coherence_temperature} and Sec. \ref{sec:spinLT}.). Even a larger spin polarization into one or two hyperfine levels accomplished by scanning the laser over a specific broad frequency range did not lead to appreciable variations in $\Gamma_h$, thereby confirming that FF processes do not represent
a significant contribution  to the homogeneous linewidth at this $^{171}$Yb concentration. 


The lack of linewidth narrowing below 2\,K also excludes the interaction with TLS from the major dephasing factors. Instantaneous spectral diffusion can be ruled out as well since we did not observe any narrowing by reducing the laser power by a factor of 10 in the 2PE experiments (SM sec.\,\ref{sec:ISD}). 

Second-order $^{171}$Yb interactions with nuclear spins or paramagnetic defects become therefore the main candidate to explain the residual linewidth broadening of 200 Hz compared to the lifetime limited width of 122 Hz. These interactions could also be increased by slight deviations from the ZEFOZ point caused by a small magnetic field background at the sample location.  Other instrumental limitations such as laser instability and crystal vibrations may also play a role as we could observe lines as narrow as $255 \pm 16$ Hz in a different set-up \cite{lafitte2022optical}. 

Measurements of 3PE decays revealed a broadening of the homogeneous linewidth on time scales of 100s of µs that increased from 200 to 400 Hz between 0.3 and 4.5 K for delays $\mathrm{T_W}$ of 1-5 ms (Fig. \ref{fig:spectral_diffusion}). This clear increase seems to exclude experimental bias such as laser instability and cryostat vibrations. The source of this broadening is however unclear as SHB measurements revealed $^{171}$Yb spin-flip rates seemingly too slow ($\ll 1$ ms$^{-1}$) to be associated with the onset of the broadening. The same conclusion can be drawn for $^{89}$Y spin flip-flops which should occur on a 10s of ms timescale \cite{Bottger2006}. This suggests that contributions from paramagnetic defects and impurities could be responsible for the observed broadening. Another possible cause might be the superhyperfine Y-Yb interaction enabled by a non-zero background magnetic field as discussed in Ref.\,\onlinecite{Nicolas2022coherent} in relation to a hyperfine transition in $^{171}$Yb:YSO. 



We now turn to the analysis of the 10 ppm sample homogeneous linewidth. By comparison with the two samples at 2\,K, we can conclude that the higher doping is responsible for an additional linewidth broadening of 680\,Hz (Fig. \ref{fig:Coherence_temperature}), leading to $\Gamma_h = 1$ kHz. 
This broadening has been ascribed to $^{171}$Yb spin FF in Ref.\cite{Welinski2020}. Regarding mechanisms, it is interesting to note that the spin lifetime due to FF measured in Sec. \ref{sec:spin_lifetime} causes only a broadening of $1/2\pi\mathrm{T_{ff}}=80\,$Hz.  Most of the 680\,Hz total concentration-dependent broadening must then originate from the indirect effect of magnetic noise created by the bath of non-optically probed $^{171}$Yb ions. 
In particular, results of  Sec. \ref{sec:spin_lifetime} suggest that $^{171}$Yb ions in site II have a faster FF rate compared to those in site I and could be mainly responsible for the optical dephasing at low temperatures. 
It was actually shown that suppression of spin flip-flops, by optically polarizing $^{171}$Yb spins in site II in the 2g and 4g levels, led to a line narrowing down to 400\,Hz \cite{Welinski2020}. Such value is separated by only 80\,Hz from the homogeneous linewidth of the 2\,ppm sample. 

Below 2\,K, a linewidth narrowing is observed for decreasing temperature and is also believed to be provoked by the reduction of the FF rate between the 4g and 3g levels, due to partial thermal spin polarization.  The narrowest measured line is however 700\,Hz broad, significantly above the value of the 2\,ppm. This may be explained by a spin temperature much larger than the base plate one, as already mentioned in Sec.\,\ref{sec:spin_lifetime} for SHB measurements.

Finally, spectral diffusion, investigated by 3PE at 1.2 K, shows a similar trend as in the 2 ppm sample. Indeed, an onset of broadening at 100s of $\mu$s is observed. It however reaches a higher value, of about 2\,kHz for $\mathrm{T_W}$ delays of a few ms. This additional broadening compared to the 2 ppm sample could be due to magnetic noise provoked by FF processes that are acting on a $\mathrm{T_{ff}}= 2 $ ms time scale.



\subsection{High temperature regime ($> 6$ K)}

Above 6\,K, the decoherence rate increases with the ninth power of the temperature in both crystals. In Sec. \ref{SpinHT}, the spin relaxation rate has been also found to scale with $T^9$, and attributed to a two-phonon Raman process. According to Eq.\,\ref{Eq:linewidth_broadening}, the SLR of the level 1e and 4g induces an optical homogeneous broadening for the 1e-4g transition of $2/(2\pi \mathrm{T_{S}})=\alpha_RT^9/\pi$, however, this contribution is only about ($\alpha_R/\pi\gamma_R)\sim24$\%  of the total homogeneous linewidth decoherence rate $\gamma_RT^9$. By difference, the remaining part still follows the $T^9$-dependence and is roughly independent of the Yb concentration. We therefore attribute it to an elastic Raman scattering, i.e. absorption and emission of phonons with the same energy so the ion energy level remains unperturbed but the phase of the superposition state is randomly changed. 
We note however that the temperature dependence of this process is usually reported as $T^7$ \cite{mccumberLinewidthTemperatureShift1963,bottgerOpticalSpectroscopyDecoherence2016}, which we also approximately found for temperatures higher than 35 K by directly measuring the absorption linewidth in the 2 and 10 ppm samples \cite{lafitte2022optical}. The unusual $T^9$ dependence for the optical dephasing could be due to the nature of hyperfine sublevels of $^{171}$Yb in YSO which are symmetrically hybridized and possess identical non-zero projections of both the electron spin-up and spin-down at zero magnetic field. In this sense, the elastic and inelastic Raman processes could be similar and the T$^9$-temperature dependence expected in both cases. We note that at high temperatures, 76\% of the dephasing rate is caused by the elastic Raman process, which is then three times larger than the total inelastic Raman contribution.


Nevertheless, as pointed out in Sec. \ref{sec:2PE}, the homogeneous linewidth broadening with temperature is quite small and comparable to Eu$^{3+}$ in YSO for example \cite{konzTemperatureConcentrationDependence2003}. This is due in part to the large energy separation, about 250 cm$^{-1}$ between the lowest crystal field levels of the ground and excited states which prevent fast electron spin relaxation, and thus coherence loss, by direct phonon absorption.  
A slightly larger coefficient $\gamma_R$ was also found for the 10\,ppm compared to the 2\,ppm one. This is attributed to the effect of the stronger spectral diffusion due to Yb spin flips in the 10 ppm sample which contribution becomes negligible only at temperatures above about 12 K, where the 2 ppm data end. As a result, the fitted $\gamma_R$ values are larger for the 10 ppm sample. 

Spectral diffusion is also present in the high-temperature regime and is attributed in both samples  to magnetic noise due to $^{171}$Yb spin flips. Indeed the spin lifetimes determined in Sec. \ref{SpinHT} are consistent with the delays at which broadening is observed, i.e. 100s of $\mu$s. Moreover, spectral diffusion is stronger in the 10 ppm, as expected for an indirect decoherence mechanism. For example, by looking at the data\ recorded for both samples at 7.5\,K (orange points in Fig. \ref{fig:spectral_diffusion}), the linewidth broadening at several ms delays is three times larger in the 10\,ppm sample than in the 2\,ppm one.

Above 9\,K no broadening due to spectral diffusion is observed anymore in both samples. The spin lifetimes become much shorter than the delay times used to estimate the linewidths and the maximum contribution of spectral diffusion is already included in the linewidths measured at the shortest waiting times probed.

\section{Conclusion}

A detailed spectroscopic study of $^{171}$Yb$^{3+}$, a promising species for quantum memories, has been performed. It focuses on ions in site II of YSO at doping levels of 2 and 10\,ppm and is aimed at elucidating contributions to optical homogeneous linewidths. All measurements were performed in the zero-field ZEFOZ point, where long optical coherence times are observed. First, $^{171}$Yb spin population dynamics has been recorded using SHB and 3PE from 40\,mK to 10\,K. A modeling of these experiments including spin relaxations in the ground and excited states as well as optical relaxation between ground and excited states hyperfine levels  was developed in low- and high-temperature regimes. In the former ($T<$5\,K), FF processes dominate ground state spin dynamics, whereas in the latter ($T>$6\,K), SLR through a two-phonon Raman process is the main relaxation mechanism. Very good agreement with experimental data was obtained, supporting the proposed modeling. This allowed us to estimate the direct $T_1$-contribution to the optical homogeneous linewidth as a function of temperature in both samples.

These predictions were then compared to optical homogeneous linewidths $\Gamma_h$ measured in the range from 40\,mK to 18\,K by 2PE and SHB.  Variations of $\Gamma_h$ were also investigated as a function of time using 3PE. The first conclusion is that direct contributions to $\Gamma_h$ either arising from spin or optical relaxation are never the predominant decoherence mechanism. In the low-temperature regime and 2\,ppm sample, dephasing is attributed to magnetic interactions with $^{89}$Y nuclear spins or other paramagnetic impurities or defects. Thanks to $^{171}$Yb ZEFOZ transitions at zero magnetic field, these interactions lead to a remarkable narrow line of 320\,Hz from 0.04 to 3\,K. In the 10\,ppm, magnetic noise from $^{171}$Yb flip-flops is identified as the main  source of decoherence and spectral diffusion. In this case, partial spin polarization at temperatures $\leq$ 2\,K is able to lower this effect. In the high-temperature regime, decoherence is attributed for both $^{171}$Yb concentrations to an elastic two-phonon process that slowly broadens lines. A 25\,kHz linewidth is measured at 10 K, which could be of interest for relaxing cryogenic requirements in applications such as RF signal analysis and medical imaging by ultra-sound optical tomography. 

Our study indicates that a further decrease of homogeneous linewidths in the 2 ppm sample below 5 K could be obtained by shielding samples from stray magnetic fields and/or decreasing concentrations of remaining paramagnetic impurities or defects. Decreasing concentrations or spin polarization at ultra-low temperatures is not relevant at this stage. In the 10 ppm sample, the latter is however found partially effective. At temperatures $>$\,6\,K, dephasing is intrinsic to Yb ions, suggesting that relatively narrow linewidths could be obtained at higher Yb concentrations.

\begin{acknowledgments}
    This project has received funding from the European Union’s Horizon 2020 research and innovation programme under Grant Agreement No. 820391 (Square), the ANR MIRESPIN project, Grant No. ANR-19-CE47-0011 of the French Agence Nationale de la Recherche, the Plan  France 2030 project ANR-22-PETQ-0010, and the Swiss National Science Foundation (SNSF) project No. 197168. The authors also acknowledge support from DGA.
\end{acknowledgments}

\appendix*

\section{Experimental methods}
\subsection{Spectral hole burning}

\begin{figure}[h!]
	\includegraphics[width=1\linewidth]{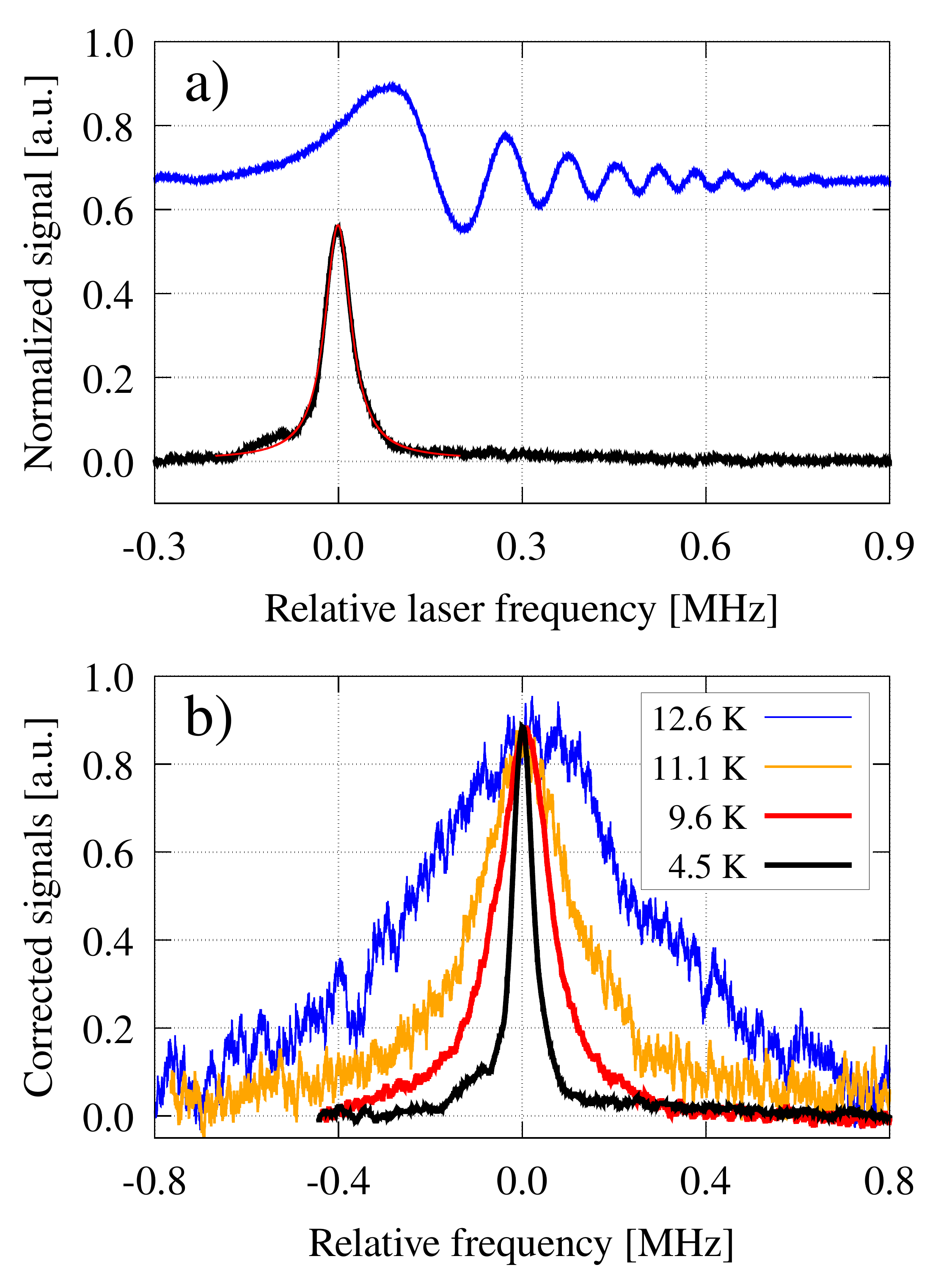}
	\caption{a) Transmission signal (in blue) and deconvoluted spectral hole profile (in black) obtained burning the 4g$\rightarrow$1e transition at 4.5\,K. The red line is a Lorentzian fit. b) Representative examples of the deconvoluted spectral holes for different temperatures.}
	\label{fig:hole_linewidth}
\end{figure}

For the homogeneous linewidth estimation, we used 25\,$\mu$s-long burning pulses with fixed frequency and power of tens of mW/cm$^2$. The created spectral feature was temporally mapped after a delay of 15\,$\mu$s to a frequency-chirp readout laser pulse. Its duration was $\mathrm{T_r}= 60\,\mu$s and the scanned frequency $\Delta \nu=2\,$MHz for the 7-10 K temperature range. With the increase of the temperature and the homogeneous linewidth, the frequency range scanned by the readout pulse was adjusted, up to 40\,MHz. The transmitted signal averaged over tens of traces at 4.5\,K is shown in blue in Fig.\,\ref{fig:hole_linewidth}a. It is evident that the spectral mapping has been distorted. Actually, whenever  $\Delta \nu/\mathrm{T_r} \gg \Gamma_{hole}^2$, the interference of the chirp laser radiation with the coherent emission of the previously excited ions introduces a beat that strongly modifies the hole profile. The phenomenon was already addressed in the literature and a data processing technique to extrapolate the right hole profile was developed\,\cite{chang2005recovery}. The correction procedure consists of adding a factor to the Fourier-transformed signal before doing the inverse Fourier transformation. The factor depends only on the laser pulse parameters $\Delta \nu$ and $\mathrm{T_r}$, so no fit-parameter is required. The logarithm of the chirp-compensated signal, corresponding to the true hole profile, is shown in black in Fig.\,\ref{fig:hole_linewidth}a. The profile is Lorentzian with a linewidth of 56.6\,kHz (see red line). The burning power set at $\sim$10$\mu$W was weak enough to make the power broadening negligible as the decrease or slight increase of the burning pulse power did not lead to a substantial change in the hole width. According to our finding in the sec.\,\ref{sec:spectral_diffusion}, the spectral diffusion during the whole sequence duration, equal to 100\,$\mu$s, should be less than 1\,kHz below 7\,K. As we measured homogeneous linewidth of 1\,kHz at 4.5\,K by 2PE, we can conclude that the hole linewidth is predominantly determined by the laser linewidth plus the laser frequency jitter. Indeed, by averaging, we can achieve a good signal-to-noise ratio but the laser frequency fluctuation occurring during $\mathrm{T_D}$  induced a random shift in the hole center position in each trace and thus the broadening of the hole linewidth in the averaged signal. By fitting a single-trace signal, compensated for the fast chirp effect, the linewidth is roughly 10\,kHz which is close to the Fourier transform  of the pulse duration. However, the use of longer pulses did not improve the hole width, indicating that the limit of the laser has been reached. Even though the hole width is narrower in the single trace than in the averaged trace, its signal is affected by a large noise and the hole width cannot be estimated accurately. The uncertainty is even more relevant at higher temperatures where the holes are less deep and broadened. For these reasons, we used the averaged signals. A few corrected hole profiles are shown in Fig.\,\ref{fig:hole_linewidth}. Above 13\,K, to record deep enough holes, we had to increase the burning pulse power up to hundreds of $\mu$W, paying attention that the power-induced broadening was less than a few percent of the other factors. The homogeneous linewidth was finally determined as $\Gamma_h = (\Gamma_\mathrm{hole}-56.6\,\mathrm{kHz})/2$.

For the hole area decay measurements, 2\,MHz-broad holes were burnt by sweeping the laser frequency during the burning pulse. As the inhomogeneous linewidth is hundreds of MHz for both samples, only a small fraction of the ions were excited from the ground state, and therefore, the spin dynamics were not altered by the burning pulse. A power between tens and hundreds of $\mu$W was used for a duration of 10\,$\mu$s. The evolution of the hole area was monitored using reading pulses delayed between a few microseconds up to milliseconds and spin relaxation as short as tens of microseconds could be investigated in this way.

\subsection{Two-pulse photon echo}

The 2PE pulse sequence used in the experiments is a 0.6\,$\mu$s-long $\pi/2$ square pulse followed by a 1.2\,$\mu$s-long $\pi$ square pulse. The pulse peak power was $\sim$10\,mW and the delay time $\tau_d$ between  the two pulses has been varied between 3 and 600\,$\mu$s. 
The mechanical vibrations of the crystal due to the refrigerator induced large fluctuations of the PE signal for $\tau_d >$50\,$\mu$s. To record the full extension of the atomic coherence, tens of echo signals were generated for each delay time and the strongest one was selected by the oscilloscope using the roof function. At temperatures below 5\,K, the spin-lattice relaxation may not be fast enough to prevent optical pumping of ground state spin levels. The resulting decrease in absorption due to the repetition of the laser pulses provokes a gradual lowering of the echo signal, impacting the measurement of the homogeneous linewidth. To mitigate this effect, the pulse repetition rate was reduced from a few Hz to hundreds of mHz along with the overall number of pulse sequences generated for the estimation of each homogeneous linewidth. In addition, the laser frequency was scanned during the repetition of the 2PE sequence by $\sim$10\,MHz at 10\,Hz. 
The remaining decrease in absorption by optical pumping was quantified and subtracted. We recorded the decrease of the 2PE signal over the repetition of several hundreds of identical sequences. The decrease could be well fitted with the function $P(n)=a+b\cdot e^{-n/c}$ where $n$ is the number of the pulse sequence. The final peak values of the PE signals obtained consecutively for different $\tau_d$ were accordingly corrected with the obtained function. This procedure has been repeated for every investigated temperature. In all cases, the correction of the peak values did not exceed 30\% of the raw value, while 2PE decays varying over more than two orders of magnitude as a function of $\tau_d$  were generally collected and fitted. 
 
 \bibliography{My_collection}
 \makeatletter\@input{yy.tex}\makeatother
\end{document}


	\title{Supplementary materials: optical coherence and spin population dynamics in $^{171}$Yb$^{3+}$:Y$_2$SiO$_5$ single crystals}
	
	\author{Federico Chiossi} 
 \email{federico.chiossi@chimieparistech.psl.eu}
 \affiliation{Chimie ParisTech, PSL University, CNRS, Institut de Recherche de Chimie, Paris, 75005, France}

\author{Eloïse Lafitte-Houssat}	
\affiliation{Chimie ParisTech, PSL University, CNRS, Institut de Recherche de Chimie, Paris, 75005, France}
 \affiliation{Thales Research and Technology, 91767 Palaiseau, France}
\author{Alban Ferrier}	
\affiliation{Chimie ParisTech, PSL University, CNRS, Institut de Recherche de Chimie, Paris, 75005, France}
\affiliation{Faculté des Sciences et Ingénierie, Sorbonne Université, UFR 933, 75005 Paris, France}

\author{Sacha Welinski}
\author{Lo\"ic Morvan}

 \affiliation{Thales Research and Technology, 91767 Palaiseau, France}

\author{Perrine Berger}
\affiliation{Thales Research and Technology, 91767 Palaiseau, France}
 
\author{Diana Serrano}
\affiliation{Chimie ParisTech, PSL University, CNRS, Institut de Recherche de Chimie, Paris, 75005, France}
\author{Mikael Afzelius}
 \affiliation{Department of Applied Physics, University of Geneva, Geneva 4, Switzerland}
 
\author{Philippe Goldner}
\email{philippe.goldner@chimieparistech.psl.eu}	
 \affiliation{Chimie ParisTech, PSL University, CNRS, Institut de Recherche de Chimie, Paris, 75005, France}

 \maketitle
	
\modulolinenumbers[2]

\section{Experimental setup}
\label{sec:setup}

\begin{figure}[h!]
	\includegraphics[width=0.5\linewidth]{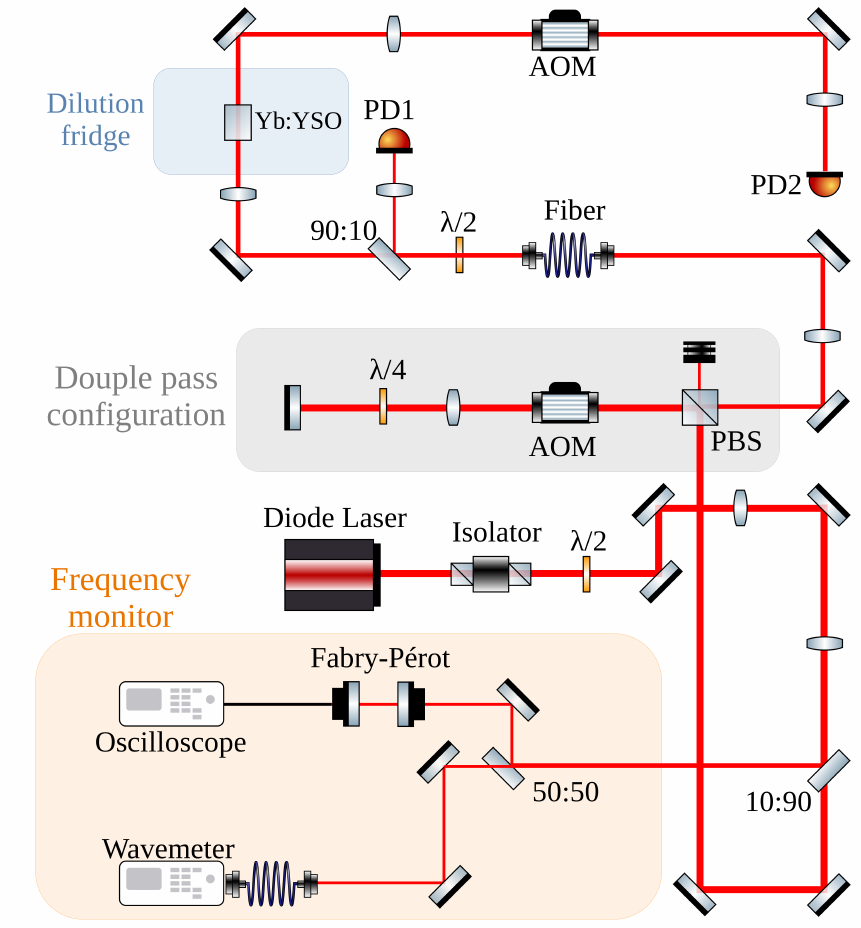}
	\caption{Experimental scheme. Legend: AOM=acousto-optic modulator. PBD=polarized beam splitter. PD=photodetector. $\lambda$/2=half-wave plate. $\lambda$/4 = quater-wave plate.  }
	\label{fig:scheme}
\end{figure}

The experimental setup is sketched in Fig.\,\ref{fig:scheme}. The laser diode Toptica DLpro is tunable between 930 and 990\,nm. Its absolute frequency is measured by a wavemeter (Burleigh WA1000) with an accuracy of 100\,MHz while the relative frequency during the laser scanning was monitored with a precision of a few MHz using a Fabry-Pérot interferometer.  The laser beam was shaped in power and manipulated in frequency to create the photon echo and the burning-readout pulse sequences for the spectral hole burning by means of an acousto-optic modulator (Keysight M3201A) in the double pass configuration, driven by an arbitrary waveform generator (AWG, Agilent N8242A). The output beam was then sent to the Yb:YSO crystal placed inside a dilution fridge (Bluefors SD) endowed with optical windows. Two photodetectors measured the laser power before (Thorlabs PDB450A - detector PD1) and after the crystal (Hamamatsu c5460 - detector PD2) in order to accurately determine the crystal absorption. In this way, the full absorption spectrum shown in Fig.\,\ref{fig:absorption} of the manuscript was recorded. PD2 is also used to record the photon echo intensity and the laser chirp pulses. A time gate for PD2 was introduced using a second identical AOM (Keysight M3201A) driven by the same arbitrary waveform generator to avoid the detector saturation because of the preparatory photon echo or burning pulses.

\section{Determination of the spin-flip rate}
\label{sec:model_sf}

Both the spectral hole area and the intensity of the 3PE depend on the initial perturbation of the population distribution of the ions and its time evolution. When several sublevels are involved, the relationship can become very complex but in our case, an analytical expression with one or two free parameters could be calculated under some reasonable assumptions. The characteristic relaxation time of the ions, i.e. the spin-flip rate, can be thus estimated by fitting the time evolution of the spectral hole area and the intensity of the 3PE as a function of the waiting time. We distinguished between two situations and we developed a model for the spin relaxation for each situation. The first model assumed that the SLR is negligible and considered only the FF processes. It is thus valid only for low temperatures. The second model required that the spin-flip at a short time scale is completely due to SLR and can thus be applied when the temperature is sufficiently high to neglect the FF processes.   

In our models, we consider only the lowest Stark levels of the ground and excited manifold. Each level is split into four non-degenerate hyperfine sublevels, labeled for increasing energy position as $\mathrm{1g}$, $\mathrm{2g}$, $\mathrm{3g}$ and $\mathrm{4g}$ for the ground state level and $\mathrm{1e}$, $\mathrm{2e}$, $\mathrm{3e}$ and $\mathrm{4e}$ for the excited level (see Fig.\,\ref{fig:absorption} of the manuscript).

\subsection{First model}
\label{sec:model2}

At sufficiently low temperatures, the spin-phonon interaction rate is much smaller than the interaction between spin-spin, and the spin flip-flop can be considered as the only process affecting the spins in the ground state. 

The holes in the absorption spectrum are thus re-absorbed due to the excited ion relaxation from the optical levels and the spin flip-flop process only. The latter process makes two ions with similar spin transition frequency exchange their spin states so that the population distribution over the different levels is not modified. However there is no correlation between optical and spin transitions, namely ions with a similar optical transitions frequency generally do not have similar spin transition frequencies, and vice versa. As a result, the occurrence of FF processes progressively reduces the depth of the holes spreading the deficit of absorbing ions from the hole to the whole inhomogeneous transition. 


To model the hole relaxation, we must discriminate the ions in the different levels that are resonant to the laser when they occupy the level 4g. This subclass will be hereafter identified by adding a bar on top of the population symbol. Note that due to the narrowness of the hole, this subclass is a negligible fraction ($\sim0.2\%$) of the total amount so the perturbation induced by the laser excitation does not modify appreciably the rate of the ground spin dynamics. As discussed in Ref.\,\cite{Welinski2020}, the FF rate strongly depends on the wavefunction of the states of the involved ions. In particular, it has been calculated a strong coupling between the 3g and 4g levels and a much weaker coupling for the 3g and 4g levels with the 1g and 2g levels. For the last cases, the FF rate is much slower than the radiative emission rate and the processes can be neglected in the timescale of a few lifetimes. 

Let us define $\mathrm{T_{ff}}$ the characteristic time for an ion in a specific crystal to accomplish a $\mathrm{3g\rightleftarrows 4g}$ FF process. The rate equation for the ion population in level 4g resonant to the laser can be written as:
\begin{equation}
\dfrac{d\overline{n}_\mathrm{4g}}{dt} = -\dfrac{\overline{n}_\mathrm{1e}}{\mathrm{T_{opt}}}\beta_{14}+\dfrac{\overline{n}_\mathrm{3g}-\overline{n}_\mathrm{4g}}{\mathrm{T_{ff}}}
\label{Eq:6} 
\end{equation}
Since there is no spin-flip in the optical levels, the population in $\mathrm{1e}$  can be expressed as:
\begin{equation*}
{\overline{n}_\mathrm{1e}} = {Ne^{-t/\mathrm{T_{opt}}}},
\end{equation*}
where N is the number excited by the burning pulse. In addition,
\begin{equation}
{\overline{n}_\mathrm{3g}} = \overline{n}_\mathrm{4g}^\mathrm{eq}+\overline{n}_\mathrm{2g}^\mathrm{eq}+\overline{n}_\mathrm{3g}^\mathrm{eq}+\overline{n}_\mathrm{4g}^\mathrm{eq} - {\overline{n}_\mathrm{4g}} - {\overline{n}_\mathrm{2g}} -{\overline{n}_\mathrm{1g}}- {\overline{n}_\mathrm{1e}} 
\label{Eq:7-}
\end{equation} 
If the temperature is not lower than 1\,K, we can expect an equal distribution of the ions over the four ground hyperfine levels. Then:
\begin{equation}
{\overline{n}_\mathrm{3g}} = 4\overline{n}_\mathrm{1g}^\mathrm{eq} - {\overline{n}_\mathrm{4g}} - {\overline{n}_\mathrm{2g}} -{\overline{n}_\mathrm{1g}}- {\overline{n}_\mathrm{1e}} 
\label{Eq:7}
\end{equation} 
As we neglect the FF involving the first two levels, their population is only altered by the relaxation from the 1e level to these two levels
\begin{equation}
{\overline{n}_\mathrm{1g}} + {\overline{n}_\mathrm{2g}} = 2\overline{n}_\mathrm{4g}^\mathrm{eq} + N\left(1-e^{-t/\mathrm{T_{opt}}} \right)(\beta_{11}+\beta_{12})
\end{equation} 

By substituting the last equation in Eq\,\ref{Eq:7}:
\begin{equation}
{\overline{n}_\mathrm{3g}} = 2\overline{n}_\mathrm{4g}^\mathrm{eq} - {\overline{n}_\mathrm{4g}} -  N\left(1-e^{-t/\mathrm{T_{opt}}}\right)(\beta_{11}+\beta_{12})- {Ne^{-t/\mathrm{T_{opt}}}}
\label{Eq:8}
\end{equation} 

The rate equation Eq\,\ref{Eq:6} can be replaced by:
\begin{equation}
\dfrac{\overline{n}_\mathrm{4g}}{dt} = {Ne^{-t/\mathrm{T_{opt}}}}\left(\frac{\beta_{14}}{\mathrm{T_{opt}}} +\frac{\beta_{11}+\beta_{12}-1}{\mathrm{T_{ff}}}\right)-\frac{2\overline{n}_\mathrm{4g}^\mathrm{eq}-2\overline{n}_\mathrm{4g}-N(\beta_{11}+\beta_{12})}{\mathrm{T_{ff}}}
\label{Eq:19} 
\end{equation}

Taking into account the boundary condition $\overline{n}_\mathrm{4g}(0) = \overline{n}_\mathrm{4g}^\mathrm{eq}-N$, the solution for the differential equation is:

\begin{equation}
\overline{n}_\mathrm{4g}^\mathrm{eq}-{\overline{n}_\mathrm{4g}} = N\left\{e^{-2t/\mathrm{T_{ff}}}+\left(e^{-t/\mathrm{T_{opt}}}-e^{-2t/\mathrm{T_{ff}}} \right) \left[\frac{(1-\beta_{11}-\beta_{12})\mathrm{T_{opt}}-\beta_{14}\mathrm{T_{ff}} }{2\mathrm{T_{ff}}-\mathrm{T_{opt}}}\right]+\frac{\beta_{12}+\beta_{11}}{2}\left(1-e^{-2t/\mathrm{T_{ff}}}\right)\right\}
\label{Eq:20} 
\end{equation}

The hole area is proportional to $\overline{n}_\mathrm{4g}(0)$ and $\mathrm{T_{ff}}$ has been estimated by fitting the hole area decay (Fig.\,\ref{fig:ff} of the manuscript). We used the branching ratios $\beta_{14}=0.72$ and $\beta_{11}+\beta_{12}=0.21$ estimated in Ref.\,\onlinecite{Welinski2020} for our analysis, which are proven to work fine. Note that the model is not valid for long timescale as the slow FF processes have not been taken into account. 

\subsection{Second model}
\label{sec:model1}

The second model is based on two assumptions: (i) the phonon-ion coupling for the ions occupying the ground level and the excited level is identical, (ii) each ion will move with the same probability from the occupied sublevels to one of the other three of the same level with a rate of 1/$\mathrm{3T_s}$. 

Both in the SHB and 3PE experiments, the laser excitation was tuned at the transition $4g\rightarrow1e$. Let be $N$ the number of ions out of $n_\mathrm{4g}^\mathrm{eq}$ that have been pumped from the level $\mathrm{4g}$ to the level $\mathrm{1e}$ at the end of the burning pulse (t=0) for the SHB experiment, or the first two $\pi/2$ pulses for the 3PE. The ions in both levels experience SLR. In addition, the ion in level $i\mathrm{e}$ can also relax to the level $j\mathrm{g}$ with a rate $\beta_\mathrm{ij}/\mathrm{T_{opt}}$, where the branching ratio $\beta_\mathrm{ij}$ were estimated from the absorption spectrum in the Supplementary materials of Ref.\,\onlinecite{Welinski2020}.

The rate equation for the level population in $\mathrm{1e}$ can be thus expressed as:
\begin{equation*}
\dfrac{dn_\mathrm{1e}}{dt} = -\dfrac{n_\mathrm{1e}}{\mathrm{T_{opt}}}-\dfrac{n_\mathrm{1e}}{\mathrm{T_{s}}}+\dfrac{n_\mathrm{2e}+n_\mathrm{3e}+n_\mathrm{4e}}{3\mathrm{T_s}} 
\end{equation*}
By adding and subtracting $n_\mathrm{1e}/\mathrm{3T_s}$ on the right side and taking into account that the sum of the population in the excited sublevels evolves as $Ne^{-t/\mathrm{T_{opt}}}$, the equation can be recast as:
\begin{equation*}
\dfrac{dn_\mathrm{1e}}{dt} = -\dfrac{n_\mathrm{1e}}{\mathrm{T_{opt}}}-\dfrac{4n_\mathrm{1e}}{\mathrm{3T_{s}}}+\dfrac{Ne^{-t/\mathrm{T_{opt}}}}{3\mathrm{T_s}} 
\end{equation*}
The equation is straightforward to solve and adding the condition $n_\mathrm{1e}(0)=N$, the dynamics of the $n_\mathrm{1e}$ population reads as:
\begin{equation}
{n_\mathrm{1e}}(t) = \dfrac{N}{4}\exp\left(-\dfrac{t}{\mathrm{T_{opt}}}\right) \left[ 3\exp\left(-\dfrac{4t}{\mathrm{3T_{s}}}\right)+1    \right]
\label{Eq:boh1}
\end{equation}

For the population in $\mathrm{4g}$, we need to consider the relaxation from the optical levels besides the SLR. The following rate equation holds:
\begin{equation*}
\dfrac{dn_\mathrm{4g}}{dt} =-\dfrac{n_\mathrm{4g}}{\mathrm{T_{s}}}+\dfrac{n_\mathrm{1g}+n_\mathrm{2g}+n_\mathrm{3g}}{3\mathrm{T_s}} + \dfrac{n_\mathrm{1e}\beta_\mathrm{14}+\sum_{i=2}^{4}n_{i\mathrm{e}}\beta_{i\mathrm{4}}}{\mathrm{T_{opt}}}
\end{equation*}

Like for $n_\mathrm{1e}$, we can add and subtract $n_\mathrm{4g}/\mathrm{3T_s}$ on the right side, and note that the sum of the population in the ground state sublevels is just the population at the thermal equilibrium $4n_\mathrm{4g}^\mathrm{eq}$ minus the population in the optical sublevels $Ne^{-t/\mathrm{T_{opt}}}$. The previous equation then becomes
\begin{equation}
\dfrac{dn_\mathrm{4g}}{dt} =-\dfrac{4n_\mathrm{4g}}{\mathrm{3T_{s}}}+\dfrac{4n_\mathrm{4g}^\mathrm{eq}+Ne^{-t/\mathrm{T_{opt}}}}{3\mathrm{T_s}} + \dfrac{n_\mathrm{1e}\beta_\mathrm{14}+\sum_{i=2}^{4}n_{i\mathrm{e}}\beta_{i\mathrm{4}}}{\mathrm{T_{opt}}}
\label{Eq:boh2}
\end{equation}

For the homogeneity of the spin-lattice relaxation and the identical initial condition, $n_{2\mathrm{e}}(t)=n_{3\mathrm{e}}(t)=n_{4\mathrm{e}}(t)$ and then $\sum_{i=2}^{4}n_{i\mathrm{e}}\beta_{i\mathrm{4}}=\overline{\beta}\sum_{i=2}^{4}n_{i\mathrm{e}}$ with $\overline{\beta} = (\beta_\mathrm{24}+\beta_\mathrm{34}+\beta_\mathrm{44})/3$. By adding and subtracting $\overline{\beta}n_\mathrm{1e}/\mathrm{T_{opt}}$ in the right side of Eq.\,\ref{Eq:boh2}, we can replace the populations of the excited sublevels with their sum $Ne^{-t/\mathrm{T_{opt}}}$:

\begin{equation}
\dfrac{dn_\mathrm{4g}}{dt} =-\dfrac{4n_\mathrm{4g}}{\mathrm{3T_{s}}}+\dfrac{4n_\mathrm{4g}^\mathrm{eq}+Ne^{-t/\mathrm{T_{opt}}}}{3\mathrm{T_s}} + \dfrac{n_\mathrm{1e}(\beta_\mathrm{14}-\overline{\beta})+\overline{\beta}Ne^{-t/\mathrm{T_{opt}}}}{\mathrm{T_{opt}}}
\label{Eq:boh3}
\end{equation}

The analytical solution for $n_\mathrm{1e}$ has already been calculated as Eq.\,\ref{Eq:boh1} and by considering the boundary condition $n_\mathrm{4g}(0)= n_\mathrm{4g}^\mathrm{eq}-N$, then we obtain:
\begin{align}
n_\mathrm{4g}-n_\mathrm{4g}^\mathrm{eq} &= -N\exp\left(-\dfrac{4t}{3\mathrm{T_s}}\right)+ \frac{3N}{4}(\beta_\mathrm{14}-\overline{\beta})\exp\left(-\dfrac{4t}{3\mathrm{T_s}}\right)\left[1-\exp\left(-\dfrac{t}{\mathrm{T_{opt}}}\right) \right] + \nonumber \\ &+
N \frac{3\mathrm{T_{opt}T_s}}{4\mathrm{T_{opt}-3\mathrm{T_s}}}\left( \frac{\beta_\mathrm{14}+3\overline{\beta}}{4\mathrm{T_{opt}}}-\frac{1}{3\mathrm{T_s}} \right)\left[\exp\left(-\dfrac{t}{\mathrm{T_{opt}}}\right)-\exp\left(-\dfrac{4t}{3\mathrm{T_s}}\right) \right]
\label{Eq:population4g}
\end{align}

Eq.\,\ref{Eq:population4g} can be directly applied to fit the decay of the hole area. In this case, the free parameter $N$ takes the role of the hole area at the end of the burning pulse. As burning pulses with identical power and duration have been used to burn the hole at different temperatures, the parameter $N$ is the same for all measurements. It can be set as a constant once it has been estimated accurately by recording a hole spectrum at a specific temperature. Hence, $\mathrm{T_s}$ is the only free parameter in the SLR rate determination using the SHB. 

Finally, the analytical solution for the three-pulses photon echo intensity (in which spectral diffusion is neglected) can be calculated as $\mathrm{I(t)}=(\mathrm{I'_0}/4)(n_\mathrm{1e}+n_\mathrm{4g}^\mathrm{eq}-n_\mathrm{4g})^2$.  By using Eq\,\ref{Eq:boh1} and Eq.\,\ref{Eq:population4g}, we obtain:

\begin{align}
\mathrm{I}(t) &= \frac{\mathrm{I'_0N^2}}{4}\left\{ \dfrac{1}{4}\exp\left(-\dfrac{t}{\mathrm{T_{opt}}}\right) \left[ 3\exp\left(-\dfrac{4t}{\mathrm{3T_{s}}}\right)+1    \right]  +\exp\left(-\dfrac{4t}{3\mathrm{T_s}}\right)- \frac{3}{4}(\beta_\mathrm{14}-\overline{\beta})\exp\left(-\dfrac{4t}{3\mathrm{T_s}}\right)\left[1-\exp\left(-\dfrac{t}{\mathrm{T_{opt}}}\right) \right] + \right. \nonumber \\ &- \left.
\frac{3\mathrm{T_{opt}T_s}}{4\mathrm{T_{opt}-3\mathrm{T_s}}}\left( \frac{\beta_\mathrm{14}+3\overline{\beta}}{4\mathrm{T_{opt}}}-\frac{1}{3\mathrm{T_s}} \right)\left[\exp\left(-\dfrac{t}{\mathrm{T_{opt}}}\right)-\exp\left(-\dfrac{4t}{3\mathrm{T_s}}\right) \right] \right\}^2
\label{Eq:suppl_3PE_model}
\end{align}

This equation was used to fit the data shown in Fig.\,\ref{fig:3PE_decay} of the manuscript taking ${\mathrm{I'_0N^2}}/{4}$ and $\mathrm{T_s}$ as free parameters. 

For this model, we must consider that the polarization of the ions in the optical states may change after the SLR process. In addition, as a first approximation, we assumed that the ions just pumped into the 1e level have a random polarization since, for most cases, the SLR rate is faster than the radiative relaxation rate. Therefore, for both Eq.\,\ref{Eq:population4g} and Eq.\,\ref{Eq:suppl_3PE_model}, we used the average of the branching ratios over the three polarization obtained in Ref.\,\onlinecite{Nicolas2022coherent}, namely $\overline{\beta}=0.153$ and $\beta_\mathrm{14}=0.54$. We should stress the fact that both the hole and 3PE decays are weakly dependent on the taken branching ratios. Indeed, by using the branching ratios related to the only D2 axis polarization or by assuming identical branching ratios for all transitions, the obtained $\mathrm{T_s}$ vary generally less than 10\%. The effect on the $\alpha_R$ parameter is even smaller, within a few percent.

\section{Istantaneous spectral diffusion measurements}
\label{sec:ISD}

\begin{figure}[h!]
	\includegraphics[width=0.5\linewidth]{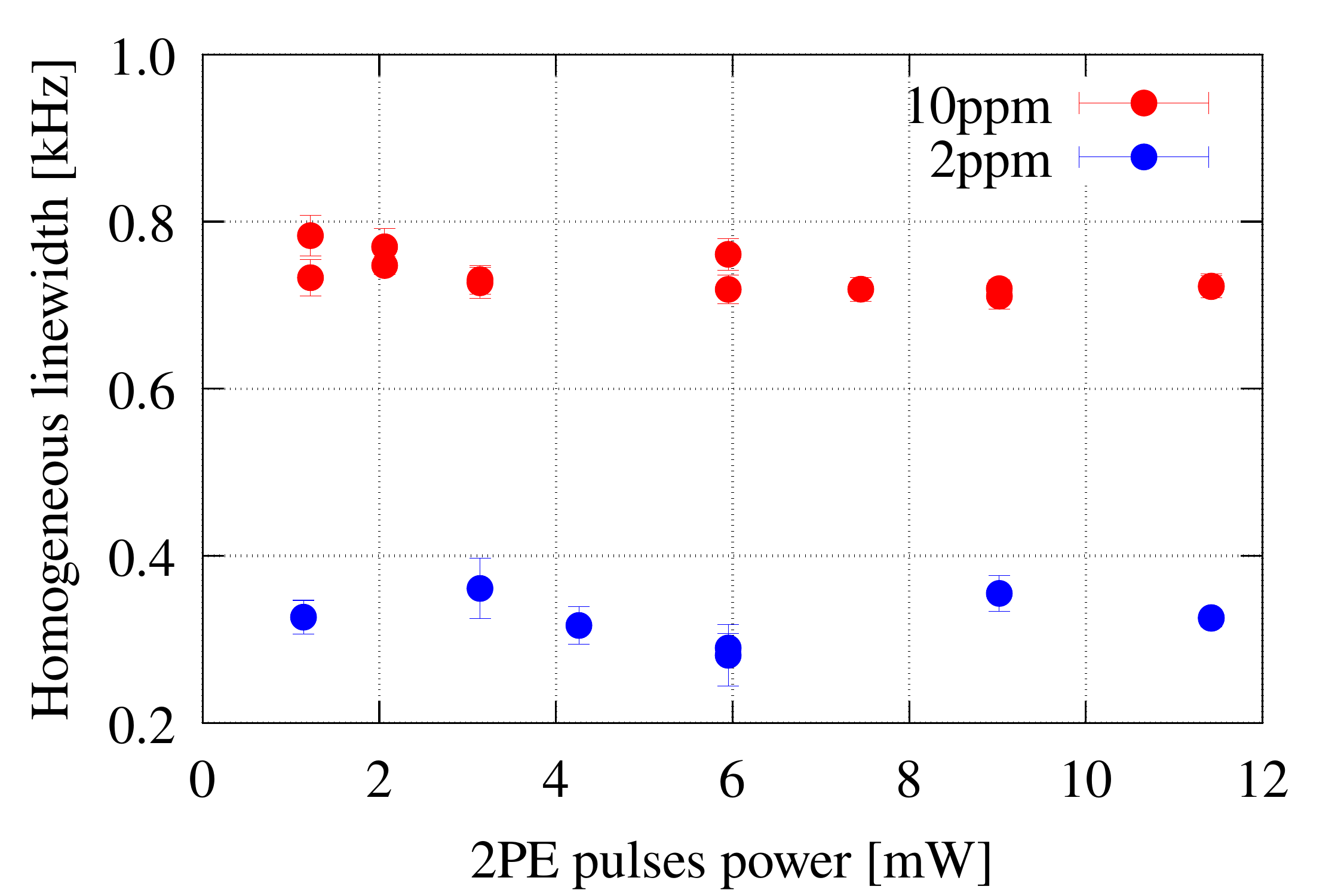}
	\caption{Homogeneous linewidth of Yb ions in the two samples estimated at 1.4\,K for different peak powers of the photon echo sequence pulses.}
	\label{fig:ISD}
\end{figure}

In this subsection, we want to study the effect of the laser pulse excitation in the 2PE experiments on the homogeneous linewidth of the excited ions. Actually, the laser excitation of other ions may significantly perturb the coherent ions. We repeated the 2PE experiments for both samples for different laser pulse peak power, starting from 11.4\,mW corresponding to the value for a $\pi/2$ pulse down to 1.1\,mW. In this way, we reduced the number of ions excited for the homogeneous linewidth measurements. The crystal temperature was set at 1.4\,K and to be completely sure that the spin distribution from one measurement to another was the same, before the 2PE sequence, we irradiated the crystal with strong laser pulses scanning the entire absorption spectrum of the 0-0 transition of site II. The obtained homogeneous linewidths are displayed in Fig.\,\ref{fig:ISD}. For both crystals, the homogeneous linewidth does not appreciably decrease with decreasing laser power, indicating that the instantaneous spectral diffusion is small. The obtained values are compatible with the one reported in Sec\,\ref{sec:coherence_time} for the 2\,ppm while they are slightly narrower ($\sim$750\,Hz vs 860\,Hz) for the 10\,ppm. This small discrepancy may be traced back to the effect of the strong laser excitation used to reset the spin distribution after each measurement. Indeed, the achieved spin distribution may differ from the thermal distribution, leading to a variation in the spin FF rate and the homogeneous linewidth.

 \bibliography{My_collection}

\makeatletter\@input{xx.tex}\makeatother